

\input harvmac       
\noblackbox

\Title{\vbox{\hbox{hep-th/9112013}\hbox{LA-UR-91-4101}}}
{Matrix Models of 2d Gravity}

\centerline{P. Ginsparg}
\bigskip\centerline{ginsparg@xxx.lanl.gov}
\smallskip\centerline{MS-B285}
\centerline{Los Alamos National Laboratory}
\centerline{Los Alamos, NM \ 87545}

\vskip .5in

\noindent
These are introductory lectures for a general audience that give an overview
of the subject of matrix models and their application to random surfaces,
2d gravity, and string theory. They are intentionally 1.5 years out~of~date.

\vskip2in\centerline
{\it Lectures given July 22--25, 1991 at Trieste summer school}

\Date{12/91}

\def\CM{{\cal M}}
\def\zbar{{\bar z}}
\def\d{{\rm d}}
\def\Sgh{S_{\rm gh}}
\def\lapl{\,\raise.5pt\hbox{$\mbox{.09}{.09}$}\,}
\def\ee{{\rm e}^}
\def\gc{g\dup_c}
\def\IR{\relax{\rm I\kern-.18em R}}
\font\cmss=cmss10 \font\cmsss=cmss10 at 7pt
\def\IZ{\relax\ifmmode\mathchoice
{\hbox{\cmss Z\kern-.4em Z}}{\hbox{\cmss Z\kern-.4em Z}}
{\lower.9pt\hbox{\cmsss Z\kern-.4em Z}}
{\lower1.2pt\hbox{\cmsss Z\kern-.4em Z}}\else{\cmss Z\kern-.4em Z}\fi}
\def\dddots{\mathinner{\mkern2mu\raise1pt\vbox{\kern7pt\hbox{.}}\mkern2mu
   \raise4pt\hbox{.}\mkern2mu\raise7pt\hbox{.}\mkern1mu}}

\ifx\epsfbox\UnDeFiNeD\message{(NO epsf.tex, FIGURES WILL BE IGNORED)}
\def\figin#1{\vskip2in}
\else\message{(FIGURES WILL BE INCLUDED)}\def\figin#1{#1}\fi

\def\ifig#1#2#3{\xdef#1{fig.~\the\figno}
\writedef{#1\leftbracket fig.\noexpand~\the\figno}%
\goodbreak\midinsert\figin{\centerline{#3}}\centerline{\vbox{\baselineskip12pt
\advance\hsize by -1truein\noindent\footnotefont{\bf Fig.~\the\figno:} #2}}
\bigskip\endinsert\global\advance\figno by1}

\lref\rDavidetal{F. David, Nucl. Phys. B257[FS14] (1985) 45, 543\semi
J. Ambj{\o}rn, B. Durhuus and J. Fr\"ohlich, Nucl. Phys. B257[FS14]
(1985) 433; J. Fr\"ohlich, in: Lecture Notes in Physics, Vol. 216,
ed. L. Garrido (Springer, Berlin, 1985)\semi
V. A. Kazakov, I. K. Kostov and A. A. Migdal, Phys. Lett. 157B (1985) 295;
D. Boulatov, V. A. Kazakov, I. K. Kostov and A. A. Migdal,
Phys. Lett. B174 (1986) 87; Nucl. Phys. B275[FS17] (1986) 641.}
\lref\rkazcon{V. Kazakov, Mod. Phys. Lett. A4 (1989) 2125.}
\lref\rBIPZ{E. Br\'ezin, C. Itzykson, G. Parisi and J.-B. Zuber,
Comm. Math. Phys. 59 (1978) 35.}
\lref\rBIZ{D. Bessis, C. Itzykson, and J.-B. Zuber,
Adv. Appl. Math. 1 (1980) 109.}
\lref\rKPZ{V. G. Knizhnik, A. M. Polyakov, and A. B. Zamolodchikov,
Mod. Phys. Lett. A3 (1988) 819.}
\lref\rDDK{F. David, Mod. Phys. Lett. A3 (1988) 1651\semi
J. Distler and H. Kawai, Nucl. Phys. B321 (1989) 509.}
\lref\rKM{V. A. Kazakov and A. A. Migdal, Nucl. Phys. B311 (1988) 171.}
\lref\rDS{M. Douglas and S. Shenker, Nucl. Phys. B335 (1990) 635.}%
\lref\rBK{E. Br\'ezin and V. Kazakov, Phys. Lett. B236 (1990) 144.}%
\lref\rGM{D. Gross and A. Migdal, Phys. Rev. Lett. 64 (1990) 127;
Nucl. Phys. B340 (1990) 333.}%
\lref\rBPZ{A. A. Belavin, A. M. Polyakov and A. B. Zamolodchikov,
Nucl. Phys. B241 (1984) 333.}
\lref\rbdss{T. Banks, M. Douglas, N. Seiberg, and S. Shenker,
Phys. Lett. B238 (1990) 279.}
\lref\rD{M. R. Douglas, Phys. Lett. B238 (1990) 176.}
\lref\rGGPZ{P. Ginsparg, M. Goulian, M. R. Plesser, and J. Zinn-Justin,
Nucl. Phys. B342 (1990) 539.}
\lref\rGZlob{P. Ginsparg and J. Zinn-Justin, Phys. Lett. B255 (1991) 189.}%
\lref\rGZaplob{P. Ginsparg and J. Zinn-Justin,
``Action principle and large order behavior of non-perturbative gravity'',
LA-UR-90-3687 / SPhT/90-140 (1990), published in proceedings of 1990
Carg\`ese workshop\semi
P. Ginsparg and J. Zinn-Justin, Phys. Lett. B255 (1991) 189.}
\lref\rkade{I. K. Kostov,  Nucl. Phys. B326 (1989) 583.}
\lref\rAGnotes{L. Alvarez-Gaum\'e, ``Random surfaces, statistical mechanics,
and string theory'', Lausanne lectures, winter 1990.}
\lref\rNotes{A. Bilal, ``2d gravity from matrix models,'' Johns Hopkins
Lectures, CERN TH5867/90\semi
V. Kazakov, ``Bosonic strings and string field theories in one-dimensional
target space,'' LPTENS 90/30, published in {\it Random surfaces and
quantum gravity\/}, proceedings of 1990 Carg\`ese workshop, edited by
O. Alvarez, E. Marinari, and P. Windey, Plenum (1991)\semi
E. Br\'ezin, ``Large $N$ limit and discretized two-dimensional quantum
gravity'', in {\it Two dimensional quantum gravity and random surfaces\/},
proceedings of Jerusalem winter school (90/91),
edited by D. Gross, T. Piran, and S. Weinberg\semi
D. Gross, ``The c=1 matrix models'', in proceedings of Jerusalem winter school
(90/91)\semi
I. Klebanov, ``String theory in two dimensions'',
Trieste lectures, spring 1991, Princeton preprint PUPT--1271
(hepth@xxx/9108019)\semi
D. Kutasov, ``Some properties of (non) critical Strings'',
Trieste lectures, spring 1991, Princeton preprint PUPT--1277
(hepth@xxx/9110041)\semi
J. Ma\~nes and Y. Lozano, ``Introduction to Nonperturbative 2d quantum
gravity'', Barcelona preprint UB-ECM-PF3/91.}
\lref\rAW{O. Alvarez and P. Windey, Nucl. Phys. B348 (1991) 490.}
\lref\rGD{I. M. Gel'fand and L. A. Dikii, Russian Math. Surveys 30:5 (1975)
77\semi
I. M. Gel'fand and L. A. Dikii, Funct. Anal. Appl. 10 (1976) 259.}
\lref\rdOne{E. Br\'ezin, V. A. Kazakov, and Al. B. Zamolodchikov,
Nucl. Phys. B338 (1990) 673\semi
G. Parisi, Phys. Lett. B238 (1990) 209, 213;
Europhys. Lett. 11 (1990) 595\semi
D. J. Gross and N. Miljkovic, Phys. Lett. B238 (1990) 217.}%
\lref\rGZ{P. Ginsparg and J. Zinn-Justin, Phys. Lett. B240 (1990) 333.}%
\lref\rdOneGK{D.J. Gross and I. Klebanov, Nucl. Phys. B344 (1990) 475;
Nucl. Phys. B354 (1991) 459.}
\lref\rFDi{F. David, Mod. Phys. Lett. A5 (1990) 1019.}
\lref\rCMM{M. L. Mehta, Comm. Math. Phys. 79 (1981) 327\semi
S. Chadha, G. Mahoux and M. L. Mehta, J. Phys. A14 (1981) 579\semi
C. Itzykson and J.B. Zuber, J. Math. Phys. 21 (1980) 411.}
\lref\rhar{Itzykson Zuber, Mehta, Harishandra, Duistermaat-Eckmann}
\lref\rWtp{E. Witten, Nucl. Phys. B340 (1990) 281.}
\lref\rKBK{V. Kazakov, Phys. Lett. 119A (1986) 140\semi
D. Boulatov and V. Kazakov, Phys. Lett. 186B (1987) 379.}
\lref\rising{E. Br\'ezin, M. Douglas, V. Kazakov, and S. Shenker, Phys. Lett.
B237 (1990) 43\semi
D. Gross and A. Migdal, Phys. Rev. Lett. 64 (1990) 717}
\lref\rIYL{C. Crnkovi\'c, P. Ginsparg, and G. Moore,
Phys. Lett. B237 (1990) 196.}
\lref\rZJ{J. Zinn-Justin, {\it Quantum Field Theory and Critical Phenomena},
Oxford Univ. Press (1989).}
\lref\rFDii{F. David, ``Phases of the large $N$ matrix model and
non-perturbative effects in 2d gravity,'' Nucl. Phys. B348 (1991) 507.}
\lref\rBMP{E. Br\'ezin, E. Marinari, and G. Parisi, Phys. Lett. B242 (1990)
35.}
\lref\rDSS{M. Douglas, N. Seiberg, and S. Shenker, Phys. Lett. B244 (1990)
381.}
\lref\rpgtr{P. Ginsparg, ``Some statistical mechanical models and conformal
field theories,'' lectures given at Trieste spring school, 1989, published in
M.  Green and A. Strominger, eds., {\it Superstrings '89}, World Scientific
1990.}
\lref\rMoore{G. Moore,
``Geometry of the string equations,'' Commun. Math. Phys. 133 (1990) 261.}
\lref\rNeu{H. Neuberger, ``Regularized string and flow equations,''
Nucl. Phys. B352 (1991) 689.}
\lref\rDIFK{P. Di Francesco and D. Kutasov, Nucl. Phys. B342 (1990) 589;
and Princeton preprint PUPT-1206 (1990) published in proceedings of Carg\`ese
workshop (1990).}
\lref\rGTW{ A. Gupta, S. Trivedi and M. Wise, Nucl. Phys. B340 (1990) 475.}
\lref\rDFK{P. Di Francesco and D. Kutasov, Phys. Lett. 261B (1991) 385\semi
P. Di Francesco and D. Kutasov, ``World sheet and space time physics in two
dimensional (super) string theory,'' Princeton preprint PUPT-1276
(hepth@xxx/9109005).}
\lref\rBerKl{M. Bershadsky and I. Klebanov, Phys. Rev. Lett. 65 (1990) 3088.}
\lref\rmdo{G. Moore, ``Double scaled field theory at $c=1$'', Rutgers
preprint RU-91-12, to appear in Nucl. Phys. B\semi
G. Moore and N. Seiberg, ``From loops to fields in 2-d quantum gravity'',
Rutgers preprint RU-91-29 (1991), to appear in Int. Jnl. Mod. Phys.}
\lref\rGKn{D. Gross and I. Klebanov, Nucl. Phys. B359 (1991) 3\semi
D. Gross and I. Klebanov, Nucl. Phys. B352 (1991) 671\semi
D. Gross, I. Klebanov, and M. Newman, Nucl. Phys. B350 (1991) 621.}
\lref\rPcbrsod{J. Polchinksi, Nucl. Phys. B346 (1990) 253.}
\lref\rColl{S. Das and A. Jevicki, Mod. Phys. Lett. A5 (1990) 1639\semi
A. Sengupta and S. Wadia, Int. Jnl. Mod. Phys. A6 (1991) 1961\semi
G. Mandal, A. Sengupta, and S. Wadia,  Mod. Phys. Lett. A6 (1991) 1465\semi
K. Demeterfi, A. Jevicki, and J.P. Rodrigues, Nucl. Phys. B362 (1991) 173,
and ``Scattering amplitudes and loop corrections in collective string field
theory. 2'', Brown preprint 803 (1991)\semi
J. Polchinski, Nucl. Phys. B362 (1991) 125.}  
\lref\rGLi{M. Goulian and M. Li, Phys. Rev. Lett. 66 (1991) 2051.}
\lref\rthooft{G. 't Hooft, Nucl. Phys. B72 (1974) 461.}
\lref\rpoly{A. M. Polyakov, Phys. Lett. 103B (1981) 207, 211.}
\lref\rO{O. Alvarez, Nucl. Phys. B216 (1983) 125.}
\lref\rMMHK{N. E. Mavromatos and J. L. Miramontes, Mod.Phys.Lett. A4 (1989)
1847\semi
E. d'Hoker and  P. S. Kurzepa, Mod.Phys.Lett. A5 (1990) 1411.}
\lref\rctbg{T. Curtright and C. B. Thorn, Phys. Rev. Lett. 48 (1982) 1309\semi
E. Braaten, T. Curtright and C. B. Thorn, Phys. Lett. 118B (1982) 115, Ann.
Phys. 147 (1983) 365\semi
E. Braaten, T. Curtright, G. Ghandour and C. B. Thorn, Phys. Rev. Lett. 51
(1983) 19, Ann. Phys. 153 (1984) 147.}
\lref\rGN{J.L. Gervais and A. Neveu, Nucl. Phys. B199 (1982) 59; B209 (1982)
125; B224 (1983) 329; B238 (1984) 125,396; Phys. Lett. 151B (1985) 271.}
\lref\rSliouv{N. Seiberg, ``Notes on quantum Liouville theory and quantum
gravity'', published in proceedings of 1990 Carg\`ese workshop\semi
J. Polchinski, ``Remarks on the Liouville field theory'',
UT Austin preprint  UTTG-19-90 (1990), presented at Strings '90 conference,
College Stn, TX\semi
E. D'Hoker, ``Continuum approaches to 2-D gravity'', UCLA/91/TEP/41,
review talk at Stonybrook Strings and Symmetries conference, May 1991.}
\lref\rFrlh{D. Friedan, Les Houches lectures summer 1982, in {\it Recent
Advances in Field Theory and Statistical Physics\/}, J.-B. Zuber and R. Stora
eds, (North Holland, 1984).}
\lref\rOA{O. Alvarez, in {\it Unified String Theories},
M. Green and D. Gross, eds., (World Scientific, Singapore, 1986).}
\lref\rGS{M. B. Green and J. Schwarz (remember them?), Phys. Lett. 149B (1984)
117.}
\lref\rpglh{P. Ginsparg, ``Applied conformal field theory'' Les Houches
Session XLIV, 1988, {\it Fields, Strings, and Critical Phenomena\/}, ed.\ by
E. Br\'ezin and J. Zinn-Justin (1989).}
\lref\rkco{I. Kostov, ``Strings embedded in Dynkin Diagrams'',
SACLAY-SPHT-90-133 (1990), published in proceedings of Carg\`ese Workshop
(1990)\semi
Phys. Lett. B266 (1991) 42.}
\lref\rtmr{S. Kharchev, A. Marshakov, A. Mironov, A. Morozov, and A. Zabrodin,
``Unification of All String Models with $c<1$'', Lebedev preprint FIAN/TD-9/91
(hepth@xxx/9111037).}
\lref\rtadt{T. Tada, Phys. Lett. B259 (1991) 442.}
\lref\rmrdt{M. Douglas, ``The two-matrix model'', published in proceedings of
1990 Carg\`ese workshop.}
\lref\rdvv{R. Dijkgraaf, H. Verlinde, and E. Verlinde, ``Notes on topological
string theory and 2D quantum gravity'', Princeton preprint PUPT-1217, published
in proceedings of Carg\`ese workshop (1990)\semi
R. Dijkgraaf, ``Topological field theory and 2d quantum gravity'',
in proceedings of Jerusalem winter school (90/91).}
\lref\rfrdc{F. David, ``Nonperturbative effects in 2D gravity and matrix
models,'' Saclay-SPHT-90-178, published in proceedings of Carg\`ese workshop
(1990).}

{\footnotefont\baselineskip12pt\ifx\answ\bigans
\noindent {0.} {Canned Diatribe, Introduction, and Apologies} \leaderfill{1}
\par
\noindent {1.} {Discretized surfaces, matrix models, and the continuum limit}
\leaderfill{3} \par
\noindent \quad{1.1.} {Discretized surfaces} \leaderfill{3} \par
\noindent \quad{1.2.} {Matrix models} \leaderfill{5} \par
\noindent \quad{1.3.} {The continuum limit} \leaderfill{9} \par
\noindent \quad{1.4.} {The double scaling limit} \leaderfill{11} \par
\noindent {2.} {All genus partition functions} \leaderfill{12} \par
\noindent \quad{2.1.} {Orthogonal polynomials} \leaderfill{12} \par
\noindent \quad{2.2.} {The genus zero partition function} \leaderfill{14} \par
\noindent \quad{2.3.} {The all genus partition function} \leaderfill{16} \par
\noindent {3.} {KdV equations and other models} \leaderfill{19} \par
\noindent \quad{3.1.} {KdV equations} \leaderfill{19} \par
\noindent \quad{3.2.} {Other models} \leaderfill{24} \par
\noindent {4.} {Quick tour of Liouville theory} \leaderfill{27} \par
\noindent \quad{4.1.} {String susceptibility $\gamma $} \leaderfill{27} \par
\noindent \quad{4.2.} {Dressed operators / dimensions of fields}
\leaderfill{33} \par
\else
\noindent {0.} {Canned Diatribe, Introduction, and Apologies} \leaderfill{1}
\par
\noindent {1.} {Discretized surfaces, matrix models, and the continuum limit}
\leaderfill{4} \par
\noindent \quad{1.1.} {Discretized surfaces} \leaderfill{4} \par
\noindent \quad{1.2.} {Matrix models} \leaderfill{6} \par
\noindent \quad{1.3.} {The continuum limit} \leaderfill{10} \par
\noindent \quad{1.4.} {The double scaling limit} \leaderfill{12} \par
\noindent {2.} {All genus partition functions} \leaderfill{13} \par
\noindent \quad{2.1.} {Orthogonal polynomials} \leaderfill{14} \par
\noindent \quad{2.2.} {The genus zero partition function} \leaderfill{16} \par
\noindent \quad{2.3.} {The all genus partition function} \leaderfill{18} \par
\noindent {3.} {KdV equations and other models} \leaderfill{21} \par
\noindent \quad{3.1.} {KdV equations} \leaderfill{21} \par
\noindent \quad{3.2.} {Other models} \leaderfill{27} \par
\noindent {4.} {Quick tour of Liouville theory} \leaderfill{31} \par
\noindent \quad{4.1.} {String susceptibility $\gamma $} \leaderfill{31} \par
\noindent \quad{4.2.} {Dressed operators / dimensions of fields}
\leaderfill{38} \par
\fi}

\bigskip
\secno-1
\newsec{Canned Diatribe, Introduction, and Apologies}

Following the discovery of spacetime anomaly cancellation in 1984 \rGS,
string theory has undergone rapid development in several directions. The
early hope of making direct contact with conventional particle physics
phenomenology has however long since dissipated, and there is as yet
no experimental program for finding even indirect manifestations of
underlying string degrees of freedom in nature \ref\rMBG{M. B. Green, private
communication (1987).}. The question of whether string theory is
``correct'' in the physical sense thus remains impossible to answer for
the foreseeable future. Particle/string theorists nonetheless continue to be
tantalized by the richness of the theory and by its natural ability to provide
a consistent microscopic underpinning for both gauge theory and gravity.

A prime obstacle to our understanding of string theory has been
an inability to penetrate beyond its perturbative expansion. Our
understanding of gauge theory is enormously enhanced by having a
fundamental formulation based on the principle of local gauge invariance
from which the perturbative expansion can be derived. Symmetry breaking and
nonperturbative effects such as instantons admit a clean and intuitive
presentation. In string theory, our lack of
a fundamental formulation is compounded by our
ignorance of the true ground state of the theory. Roughly two years
ago, there was some progress \refs{\rDS\rBK{--}\rGM} towards extracting
such nonperturbative information from string theory, at least in some simple
contexts. The aim of these lectures is to provide the conceptual
background for this work, and to describe some of its immediate consequences.

In string theory we wish to perform an integral over two dimensional geometries
and a sum over two dimensional topologies,
$$Z\sim\sum_{\rm topologies} \int\CD g\,\CD X\ \e{-S}\ ,$$
where the spacetime physics (in the case of the bosonic string) resides in the
conformally invariant action
$$S\propto\int \d^2\xi\,\sqrt g\, g^{ab}\,\del_a X^\mu\,\del_b X^\nu
\,G_{\mu \nu}(X)\ .$$
Here $\mu,\nu$ run from $1,\ldots,D$ where $D$ is the number of spacetime
dimensions, $G_{\mu \nu}(X)$ is the spacetime metric,
and the integral $\CD g$ is over worldsheet metrics. Typically we ``gauge-fix''
the worldsheet metric to $g\dup_{ab}=\ee{\ph}\delta_{ab}$, where $\ph$ is known
as the Liouville field.
Following the formulation of string theory in this form (and in particular
following the appearance of \rpoly), there was much work to develop the quantum
Liouville theory (some of which is reviewed in section 4 here), and
conformal field theory itself has been characterized as ``an unsuccessful
attempt to solve the Liouville theory'' \ref\rPolne{A. M. Polyakov, lecture at
Northeastern Univ., spring 1990.}. Evaluation of the partition function $Z$
above without taking into account the integral over geometry, however, does not
solve the problem of interest, and moreover does not provide a systematic basis
for a perturbation series in any known parameter.

The basic idea of \refs{\rDS{--}\rGM}\ relied on a discretization of the string
worldsheet to provide a method of taking the continuum limit which incorporated
simultaneously the contribution of 2d surfaces with any number of handles.
At one fowl swoop, it was thus possible not only to integrate over all possible
deformations of a given genus surface (the analog of the integral over Feynman
parameters for a given loop diagram), but also to sum over all genus
(the analog of the sum over all loop diagrams).
This would in principle free us from the mathematically fascinating but
physically irrelevant problems of calculating conformal field theory
correlation functions on surfaces of fixed genus with fixed moduli (objects
which we never knew how to integrate over moduli or sum over genus anyway).
The progress, however, is limited in the sense that these methods only apply
currently for non-critical strings embedded in dimensions $D\le1$ (or critical
strings embedded in $D\le2$), and the nonperturbative information even in this
restricted context has proven incomplete. Due to familiar problems with lattice
realizations of supersymmetry and chiral fermions, these methods have also
resisted extension to the supersymmetric case.

The developments we shall describe here nonetheless provide at least a
half-step in the correct direction, if only to organize the perturbative
expansion in a most concise way. They have also prompted much useful evolution
of related continuum methods. Our point of view here is that string theories
embedded in $D\le1$ dimensions provide a simple context for testing ideas and
methods of calculation.
Just as we would encounter much difficulty calculating infinite dimensional
functional integrals without some prior experience with their finite
dimensional analogs \ref\rSSpc{S. H. Shenker, private communication (1989).},
progress in string theory should be aided by experimentation with systems
possessing a restricted number of degrees of freedom.

These notes have been confined in content essentially to the four
lectures actually given, in order to keep them reasonably short and accessible.
(Other review references on the same general subject are
\refs{\rAGnotes,\rNotes}). This means that we stop well short
of some of the more interesting recent developments in the field
(some of which were covered by later lecturers at this school), including the
application of the critical $D=2$ dimensional models to address issues of
principle such as topology change in 2d quantum gravity, and their
relation as well to recent work on $D=2$ black holes in string theory.
We shall present no formal conclusions here other than to note that the subject
remains in active development, and we have tried at various points in the
text to draw attention to issues in need of further understanding.

\newsec{Discretized surfaces, matrix models, and the continuum limit}

\subsec{Discretized surfaces}

We begin by considering a ``$D=0$ dimensional string theory'', i.e.\ a pure
theory of surfaces with no coupling to additional ``matter'' degrees of freedom
on the string worldsheet. This is equivalent to the propagation of strings in a
non-existent embedding space. For partition function we take
\eqn\eZdo{Z=\sum_h\int\CD g\ \e{-\beta A + \gamma \chi}\ ,}
where the sum over topologies is represented by the summation over $h$, the
number of handles of the surface, and the action consists of couplings to the
area $A=\int\sqrt g$, and to the Euler character
$\chi={1\over4\pi}\int\sqrt g\,R=2-2h$.

The integral $\int\CD g$ over the metric on the surface in \eZdo\ is
difficult to calculate in general. The most progress in the continuum has
been made via the Liouville approach which we briefly review in section 4.
If we discretize the surface, on the other hand, it turns out that \eZdo\
is much easier to calculate, even before removing the finite cutoff. We
consider in particular a ``random triangulation'' of the surface
\rDavidetal, in which the surface is constructed from triangles, as in
fig.~1. The triangles are designated to be equilateral,\foot{We point out
that this constitutes a basic difference from the Regge calculus, in which
the link lengths are geometric degrees of freedom. Here the geometry is
encoded entirely into the coordination numbers of the vertices.
This restriction of degrees of freedom roughly corresponds to fixing a
coordinate gauge, hence we integrate only over the gauge-invariant moduli
of the surfaces.} so that
there is negative (positive) curvature at vertices $i$ where the number
$N_i$ of incident triangles is more (less) than six, and zero curvature
when $N_i=6$. The summation over all such random triangulations is thus
the discrete analog to the integral $\int \CD g$ over all possible geometries,
\eqn\ediscr{\sum_{{\rm genus}\ h}\ \int \CD g \quad
\to \ \sum_{\scriptstyle\rm random \atop \scriptstyle\rm triangulations}\ .}

\ifig\frandtri{A piece of a random triangulation of a surface. 
Each of the triangular faces is dual to a three point vertex of a quantum 
mechanical matrix model.}{\epsfxsize3.25in\epsfbox{randtri.ps}}

The discrete counterpart to the
infinitesimal volume element $\sqrt g$ is $\sigma_i=N_i/3$,
so that the total area $|S|=\sum_i \sigma_i$ just counts the total number of
triangles, each designated to have unit area. (The factor
of $1/3$ in the definition of $\sigma_i$ is because
each triangle has three vertices and is counted three times.)
The discrete counterpart to the
Ricci scalar $R$ at vertex $i$ is $R_i=2\pi(6-N_i)/ N_i$, so that
$$\int\sqrt g\,R\to \sum_i 4 \pi(1-N_i/6)
=4 \pi(V-\half F)=4 \pi(V-E+F)=4 \pi \chi\ .$$
Here we have used the simplicial definition which gives the Euler
character $\chi$ in terms of the total number of vertices, edges, and
faces $V$, $E$, and $F$ of the triangulation (and we have used the
relation $3F=2E$ obeyed by triangulations of surfaces, since each
face has three edges each of which is shared by two faces).

In the above, triangles do not play an essential role and may be replaced
by any set of polygons. General random polygonifications of surfaces
with appropriate fine tuning of couplings may, as we shall see, have more
general critical behavior , but can in particular always reproduce the
pure gravity behavior of triangulations in the continuum limit.

\subsec{Matrix models}

We now demonstrate how the integral over geometry in \eZdo\ may be
performed in its discretized form as a sum over random triangulations. The
trick is to use a certain matrix integral as a generating functional for
random triangulations. The essential idea goes back to work \rthooft\ on
the large $N$ limit of QCD, followed by work on the saddle point approximation
\rBIPZ.

We first recall the (Feynman) diagrammatic expansion of the (0-dimensional)
field theory integral\foot{We apologize for this recapitulation of standard
Feynman diagram technology, but prefer to keep these notes at least marginally
accessible to the mathematics community.}
\eqn\esft{\int_{-\infty}^\infty
{\d\ph\over\sqrt{2\pi}}\, \e{-\ph^2/2+ \lambda\ph^4/4!}\ ,}
where $\ph$ is an ordinary real number.\foot{The integral is understood to
be defined by analytic continuation to negative $\lambda$.} In a formal
perturbation series in $\lambda$, we would need to evaluate integrals such as
\eqn\epel{{\lambda^n\over n!}
\int_\ph\e{-\ph^2/2}\left({\ph^4\over4!}\right)^n\ .}
Up to overall normalization we can write
\eqn\essft{\int_\ph \e{-\ph^2/2}\ph^{2k}=
\left.{\del^{2k}\over\del J^{2k}}\int_\ph\e{-\ph^2/2+J\ph}\right|_{J=0}
=\left.{\del^{2k}\over\del J^{2k}}\,\e{J^2/2}\right|_{J=0}\ .}
Since ${\del\over\del J}\ee{J^2/2}=J \ee{J^2/2}$, applications of
$\del/\del J$ in the above need to be paired so that any factors of $J$
are removed before finally setting $J=0$. Therefore if we represent each
``vertex'' $\lambda\ph^4$ diagrammatically as a point with four emerging
lines (see fig.~2b), then \epel\ simply counts the number of ways to group
such objects in pairs. Diagrammatically we represent the possible pairings
by connecting lines between paired vertices. The connecting line is known
as the propagator $\langle\ph\,\ph\rangle$ (see fig.~2a) and the
diagrammatic rule we have described for connecting vertices in pairs is
known in field theory as the Wick expansion.

$$\vbox{\hrule height .7pt width 30pt\vskip30pt
\hbox{\quad(a)}}\qquad\qquad\qquad
\vbox{\hbox{\hskip20pt\vrule width .7pt height 40pt}\vskip-20pt\hrule width
40pt height .7pt\vskip30pt\hbox{\quad\ (b)}}$$
\vglue5pt\nobreak
\centerline{\footnotefont{\bf Fig.~2:}
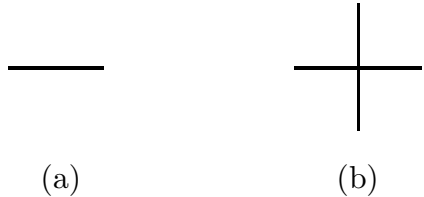
(a) the scalar propagator.   (b) the scalar four-point vertex.}
\bigbreak

When the number of vertices $n$ becomes large, the allowed diagrams begin
to form a mesh reminiscent of a 2-dimensional surface. Such diagrams
do not yet have enough structure to specify a Riemann surface. The
additional structure is given by widening the propagators to ribbons (to
give so-called ``thick'' graphs). From the standpoint of \esft,
the required extra structure is given by replacing the scalar
$\ph$ by an $N\times N$ hermitian matrix $M^i{}_j$. The analog of \essft\
is given by adding indices and traces:
\eqn\esmft{\eqalign{\int_M\e{-\tr M^2/2} M^{i_1}{}_{j_1}\cdots M^{i_n}{}_{j_n}
&=\left.{\del\over\del J^{j_1}{}_{i_1}}
\cdots{\del\over\del J^{j_n}{}_{i_n}} \e{-\tr M^2/2+\tr J M}\right|_{J=0}\cr
&=\left.{\del\over\del J^{j_1}{}_{i_1}}
\cdots{\del\over\del J^{j_n}{}_{i_n}}\, \e{\tr J^2/2}\right|_{J=0}\ ,\cr}}
where the source $J^i{}_j$ is as well now a matrix. The measure in \esmft\ is
the invariant
$\d M=\prod_i\d M^i{}_i\,\prod_{i<j}\d{\rm Re} M^i{}_j\,\d{\rm Im} M^i{}_j$,
and the normalization is such that $\int_M \ee{-\tr M^2/2}=1$.
To calculate a quantity such as
\eqn\empel{{\lambda^n\over n!}\int_M \e{-\tr M^2/2}(\tr M^4)^n\ ,}
we again lay down $n$ vertices (now of the type depicted in fig.~3b), and
connect the legs with propagators $\langle
M^i{}_j\,M^k{}_l\rangle=\delta^i_l\,\delta^k_j$ (fig.~3a).
The presence of upper and lower matrix indices is represented in fig.~3 by the
double lines\foot{This is the same notation employed in the large $N$ expansion
of QCD \rthooft.} and it is understood that the sense of the arrows is to be
preserved when linking together vertices. The resulting diagrams are similar to
those of the scalar theory, except that each external line has an associated
index $i$, and each internal closed line corresponds to a summation over an
index $j=1,\ldots,N$. The ``thickened'' structure is now sufficient to
associate a Riemann surface to each diagram, because the closed internal loops
uniquely specify locations and orientations of faces.

\font\bigarrfont=cmsy10 scaled\magstep 3
\def\extarr{\mathord-\mkern-6mu}
\def\vuline{\raise7.5pt\hbox{\textfont2\bigarrfont$\uparrow$}\hskip-4.75pt
\hbox{\vrule width.7pt depth -18 pt height 27pt}}
\def\vdline{\raise15pt\hbox{\textfont2\bigarrfont$\downarrow$}\hskip-4.75pt
\vrule width.7pt depth -4 pt height 27pt}
$$\hbox{\vbox{\hbox{\textfont2\bigarrfont$\extarr
\mathord\rightarrow\mkern-6mu\extarr\extarr$}
\vskip-10pt\hbox{$\textfont2\bigarrfont\extarr\extarr
\mathord\leftarrow\mkern-6mu\extarr$}\vskip21pt\hbox{\quad\ (a)}}
\qquad\qquad\qquad
\vbox{\hbox{$\textfont2\bigarrfont\mathord\rightarrow\mkern-6mu\extarr
\vuline\hskip2.5pt\vdline\mkern-1mu\mathord\rightarrow\mkern-6mu\extarr$}
\vskip-10pt
\hbox{$\textfont2\bigarrfont\extarr\mathord\leftarrow\mkern-6mu
\lower23pt\hbox{\vuline}\hskip2.5pt
\lower23pt\hbox{\vdline}\mkern-2mu\extarr\mathord\leftarrow$}
\vskip7pt\hbox{$\,$\qquad(b)}}}$$
\nobreak
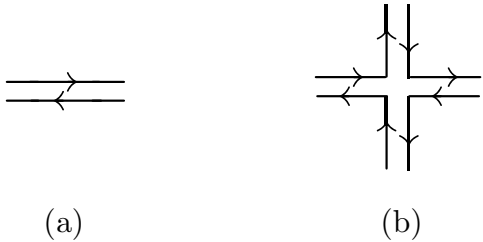
\centerline{\footnotefont{\bf Fig.~3:} (a) the hermitian matrix propagator.
(b) the hermitian matrix four-point vertex.}
\bigbreak

To make contact with the random triangulations discussed earlier,
we consider the diagrammatic expansion of the matrix integral
\eqn\ecmm{\e{Z}=\int\d M\ \e{-\ha\tr M^2+{g\over\sqrt N}\tr M^3}}
(with $M$ an $N\times N$ hermitian matrix, and the integral again
understood to be defined by analytic continuation in the coupling $g$.)
The term of order $g^n$ in a power series expansion counts the number of
diagrams constructed with $n$ 3-point vertices. The dual to such a diagram (in
which each face, edge, and vertex is associated respectively to a dual
vertex, edge, and face) is identically a random triangulation inscribed on
some orientable Riemann surface (fig.~1). We see that the matrix integral
\ecmm\ automatically generates all such random triangulations.\foot{Had we used
real symmetric matrices rather than the hermitian matrices $M$, the two indices
would be indistinguishable and there would be no arrows in the propagators and
vertices of fig.~3. Such orientationless vertices and propagators generate an
ensemble of both orientable and non-orientable surfaces.}
Since each triangle has unit area, the area of the
surface is just $n$. We can thus make formal identification with \eZdo\
by setting $g=\ee{-\beta}$. Actually the matrix integral generates both
connected and disconnected surfaces, so we have written $\ee Z$ on the
left hand side of \ecmm. As familiar from field theory, the exponential of
the connected diagrams generates all diagrams, so $Z$ as defined above
represents contributions only from connected surfaces. We see that the
{\it free energy\/} from the matrix model point of view is actually the
{\it partition function\/} $Z$ from the 2d gravity point of view.

There is additional information contained in $N$, the size of the matrix.
If we change variables $M\to M\sqrt N$ in \ecmm, the matrix action becomes
$N\,\tr(-\half\tr M^2+g\tr M^3)$, with an overall factor of
$N$.\foot{Although we could as well rescale $M\to M/g$ to pull out an overall
factor of $N/g^2$, note that
$N$ remains distinguished from the coupling $g$ in the model since it enters
as well into the traces via the $N\times N$ size of the matrix.} This
normalization makes it easy to count the power of $N$ associated to any
diagram. Each vertex contributes a factor of $N$, each propagator (edge)
contributes a factor of $N\inv$ (because the propagator is the inverse of
the quadratic term), and each closed loop (face) contributes a factor of
$N$ due to the associated index summation.
Thus each diagram has an overall factor
\eqn\efN{N^{V-E+F}=N^\chi=N^{2-2h}\ ,}
where $\chi$ is the Euler character of the surface associated to the diagram.
We observe that the value $N=\ee{\gamma}$ makes contact with the coupling
$\gamma$ in \eZdo.  In conclusion, if we take
$g=\ee{-\beta}$ and $N=\ee{\gamma}$, we can formally identify the continuum
limit of the partition function $Z$ in \ecmm\ with the $Z$ defined in \eZdo.
The metric for the discretized formulation is not smooth, but one can imagine
how an effective metric on larger scales could arise after averaging over
local irregularities. In the next subsection, we shall see explicitly how this
works.

(Actually \ecmm\ automatically calculates \eZdo\ with the measure factor
in \ediscr\ corrected to $\sum_S {1\over|G(S)|}$, where $|G(S)|$ is the
order of the (discrete) group of symmetries of the triangulation $S$. This
is familiar from field theory where diagrams with symmetry result in an
incomplete cancellation of $1/n!$'s such as in \epel\ and \empel. The
symmetry group $G(S)$ is the discrete analog of the isometry group of a
continuum manifold.)

The graphical expansion of \ecmm\ enumerates graphs as shown in fig.~1, where
the triangular faces that constitute the random triangulation are dual to the
3-point vertices. Had we instead used 4-point vertices as in fig.~3b, then the
dual surface would have square faces (a ``random squarulation'' of the
surface), and higher point vertices $(g_k/N^{k/2-1})\tr M^k$ in the matrix
model would result in more general ``random polygonulations'' of surfaces.
(The powers of $N$ associated with the couplings are chosen so that the
rescaling $M\to M\sqrt N$ results in an overall factor of $N$ multiplying the
action. The argument leading to \efN\ thus remains valid, and the power of
$N$ continues to measure the Euler character of a surface constructed from
arbitrary polygons.) The different possibilities for generating vertices
constitute additional degrees of freedom that can be realized as the coupling
of 2d gravity to different varieties of matter in the continuum limit.

\subsec{The continuum limit}

{}From \efN, it follows that we may expand $Z$ in powers of $N$,
\eqn\elne{Z(g)
=N^2 Z_0(g)+Z_1(g)+N^{-2}Z_2(g) + \ldots=\sum N^{2-2h} Z_h(g)\ ,}
where $Z_h$ gives the contribution from surfaces of genus $h$.
In the conventional large $N$ limit, we take $N\to\infty$ and only $Z_0$,
the planar surface (genus zero) contribution, survives. $Z_0$ itself may be
expanded in a perturbation series in the coupling $g$, and for large order $n$
behaves
as (see \rBIZ\ for a review)
\eqn\eloln{Z_0(g)\sim \sum_n n^{\gamma-3} (g/\gc)^n\sim (\gc-g)^{2-\gamma}\ .}
These series thus have the property that they diverge as $g$ approaches
some critical coupling $\gc$. We can extract the continuum limit of these
surfaces by tuning $g\to\gc$. This is because the expectation value of the area
of a surface is given by
$$\langle A\rangle=\langle n\rangle
={\del\over\del g}\ln Z_0(g)\sim {1\over g-\gc}$$
(recall that the area is proportional to the number of vertices $n$, which
appears as the power of the coupling in the factor $g^n$ associated to each
graph).
As $g\to\gc$, we see that $A\to\infty$ so that we may rescale the area of
the individual triangles to zero, thus giving a continuum surface with finite
area. Intuitively, by tuning the coupling to the point where the
perturbation series diverges the integral becomes dominated by diagrams
with infinite numbers of vertices, and this is precisely what we need to
define continuum surfaces.

There is no direct proof as yet that this procedure for defining continuum
surfaces is ``correct'', i.e.\ that it coincides with the continuum
definition \eZdo. We are able, however, to compare properties of the
partition function and correlation functions calculated by matrix model
methods with those (few) properties that can be calculated directly in the
continuum (for a review, see \rKM). This gives implicit confirmation that
the matrix model approach is sensible and gives reason to believe other
results derivable by matrix model techniques (e.g.\ for higher genus) that
are not obtainable at all by continuum methods.

One of the properties of these models derivable via the continuum Liouville
approach is a ``critical exponent'' $\gamma$,
defined in terms of the area dependence of the partition function for surfaces
of fixed large area $A$ as
\eqn\elpoa{Z(A)\sim A^{(\gamma-2)\chi/2-1}\ .}
To anticipate some relevant results, we recall that the
unitary discrete series of conformal field theories is labelled by an integer
$m\ge2$ and has central charge $D=1-6/m(m+1)$
(for a review, see e.g.\ \rpglh), where the central charge is normalized such
that $D=1$ corresponds to a single free boson. If we couple conformal field
theories with these fractional values of $D$ to 2d gravity, the continuum
Liouville theory prediction for the exponent $\gamma$ is (see section 4)
\eqn\epbl{\gamma={1\over12}\bigl(D-1-\sqrt{(D-1)(D-25)}\,\bigr)=-{1\over m}\ .}
The case $m=2$, for example, corresponds to $D=0$ and hence $\gamma=-\ha$ for
pure gravity. The next case $m=3$ corresponds to $D=1/2$, i.e.\ to a
1/2--boson or fermion. This is the conformal field theory of the critical Ising
model, and we learn from \epbl\ that the Ising model coupled to 2d gravity has
$\gamma=-{1\over3}$. Notice that \epbl\ ceases to be sensible for $D>1$. This
is the first indication of a ``barrier'' at $D=1$ which will reappear in
various guises in what follows.

In section 2 we shall present the solution to the matrix model formulation of
the problem, and the value of the exponent $\gamma$ provides a coarse means of
determining which specific continuum model results from taking the continuum
limit of a particular matrix model. Indeed the coincidence of $\gamma$ and
other scaling exponents (to be defined in section 4) calculated from the two
points of view were originally the only evidence that the continuum limit of
matrix models was a suitable definition for the continuum problem of interest.
In the past year, the simplicity of matrix model results for correlation
functions has spurred a rapid evolution of continuum Liouville technology so
that as well many correlation functions can be computed in both approaches and
are found to coincide.\foot{By way of very superficial overview:
following the confirmation that the matrix
model approach reproduced the scaling results of \rKPZ, some 3-point couplings
for order parameters at genus zero were calculated in \rkade\ from the
standpoint of ADE face models on fluctuating lattices. The connection to KdV
(reviewed in section 3 here) was made in \rD, and then
general correlations of order parameters (not yet known in the continuum)
were calculated in \rDIFK. The first step in the calculation of continuum
correlators was provided in \rGTW, where the free field formulation by
zero mode integration of the Liouville field was established. This was
employed in \rGLi\ together with a necessary analytic continuation of the
scaling parameter to calculate some continuum correlation functions:
the incorporation of the Liouville mode was shown to cancel the
ghastly assemblage of $\Gamma$-functions familiar from the conformal field
theory result and reproduce the relatively simple matrix model result.
Additional genus zero correlation functions for $D\le1$ were then computed
in \rDFK.
The genus one partition function for the AD series was calculated
via KdV methods in \rDIFK, and was confirmed from the continuum
Liouville approach in \rBerKl.
For $D=1$, the matrix model approach of \refs{\rdOne,\rGZ}\ was used in
\rmdo\ (also \refs{\rGKn,\rkco}) to calculate a variety of correlation
functions. These were also calculated in the collective field approach
\rColl\ where up to 6-point amplitudes were derived, and found to be in
agreement with the Liouville results of \rDFK.}

\subsec{The double scaling limit}

Thus far we have discussed the naive $N\to\infty$ limit which retains only
planar surfaces. It turns out that the successive coefficient functions
$Z_h(g)$ in \elne\ as well diverge at the same critical value of the
coupling $g=\gc$ (this should not be surprising since the divergence of the
perturbation series is a local phenomenon and should not depend on global
properties such as the effective genus of a diagram).
As we shall see in the next section, for the higher genus contributions
\eloln\ is generalized to
\eqn\elolnhg{Z_h(g)\sim \sum_n n^{(\gamma-2)\chi/2-1} (g/\gc)^n\sim
(\gc-g)^{(2-\gamma)\chi/2}\ .}
We see that the contributions from higher genus ($\chi<0$)
are enhanced as $g\to\gc$.
This suggests that if we take the limits $N\to\infty$ and $g\to\gc$ not
independently, but together in a correlated manner, we may compensate the
large $N$ high genus suppression with a $g\to\gc$ enhancement. This would
result in a coherent contribution from all genus surfaces \refs{\rDS{--}\rGM}.

To see how this works explicitly,
we write the leading singular piece of the $Z_h(g)$ as
$$Z_h(g)\sim f_h(g-\gc)^{(2-\gamma)\chi/2}\ .$$
Then in terms of
\eqn\ekappa{\kappa\inv\equiv N(g-\gc)^{(2-\gamma)/2}\ ,}
the expansion \elne\ can be rewritten\foot{Strictly speaking the first two
terms here have additional non-universal pieces that need to be subtracted
off.}
\eqn\elner{Z=\kappa^{-2}f_0+f_1+\kappa^2 f_2+\ldots
=\sum_h \kappa^{2h-2}\,f_h\ .}
The desired result is thus obtained by taking the limits $N\to\infty$,
$g\to\gc$ while holding fixed the ``renormalized'' string coupling $\kappa$
of \ekappa. This is known as the ``double scaling limit''.

\newsec{All genus partition functions}

The large $N$ limit of the matrix models considered here was originally solved
by saddle point methods in \rBIPZ. In this section we shall instead present the
orthogonal polynomial solution to the problem (\rBIZ\ and references therein)
since it extends readily to
subleading order in $N$ (higher genus corrections).

\subsec{Orthogonal polynomials}

In order to justify the claims made at the end of the previous section,
we introduce some formalism to solve the matrix models.
We begin by rewriting the partition function \ecmm\ in the form
\eqn\egpf{\e{Z}=\int \d M\ \e{-\tr V(M)}=\int\prod_{i=1}^N
\d \lambda_i\, \Delta^2(\lambda)\ \e{-\sum_i V(\lambda_i)}\ ,}
where we now allow a general polynomial potential $V(M)$. In \egpf, the
$\lambda_i$'s are the $N$ eigenvalues of the hermitian matrix $M$, and
\eqn\eVand{\Delta(\lambda)=\prod_{i<j}(\lambda_j-\lambda_i)}
is the Vandermonde determinant.\foot{\egpf\ may be derived via the usual
Fadeev-Popov method: Let $U_0$ be the unitary matrix such that
$M=U_0^\dagger \Lambda' U_0$, where $\Lambda'$ is a diagonal matrix with
eigenvalues $\lambda'_i$. The right hand side of \egpf\ follows by
substituting the definition
$1=\int \prod_i\d \lambda_i\,\d U\,\delta(U M U^\dagger-\Lambda)
\,\Delta^2(\lambda)$ (where $\int\d U\equiv1$).
We first perform the integration over $M$, and then $U$ decouples due to
the cyclic invariance of the trace so the integration over $U$ is trivial,
leaving only the integral over the eigenvalues $\lambda_i$ of $\Lambda$.
To determine $\Delta(\lambda)$, we note that only the infinitesimal
neighborhood $U=(1+T)U_0$ contributes to the $U$ integration, so that
$$1=\int \prod_{i=1}^N\d \lambda_i\,\d U\,
\delta^{N^2}\!\bigl(U M U^\dagger-\Lambda\bigr)\,\Delta^2(\lambda)
=\int\d T\ \delta^{N(N-1)}\bigl([T,\Lambda']\bigr)\Delta^2(\lambda')\ .$$
Now $[T,\Lambda']_{ij}=T_{ij}(\lambda'_j-\lambda'_i)$, so \eVand\ follows
(up to a sign) since the integration $\d T$ above is over
real and imaginary parts of the off-diagonal $T_{ij}$'s.}
Due to antisymmetry in interchange of any two eigenvalues, \eVand\ can be
written $\Delta(\lambda)=\det\,\lambda^{j-1}_i$
(where the normalization is determined by comparing leading terms).
In the case $N=3$ for example we have
$$(\lambda_3-\lambda_2)(\lambda_2-\lambda_1)(\lambda_3-\lambda_1)=
\det\pmatrix{1&\lambda_1&\lambda_1^2\cr
1&\lambda_2&\lambda_2^2\cr
1&\lambda_3&\lambda_3^2\cr}\ .$$

The now-standard method for solving \egpf\ makes use of an infinite set of
polynomials $P_n(\lambda)$, orthogonal with respect to the measure
\eqn\eopoly{\int_{-\infty}^\infty
\d \lambda\ \ee{-V(\lambda)}\,P_n(\lambda)\,P_m(\lambda)
=h_n\,\delta_{nm}\ .}
The $P_n$'s are known as orthogonal polynomials and are functions of a single
real variable $\lambda$. Their normalization is given by having leading term
$P_n(\lambda)=\lambda^n+\ldots$, hence the constant $h_n$ on the r.h.s.\ of
\eopoly. Due to the relation
\eqn\evand{\Delta(\lambda)=\det\,\lambda^{j-1}_i=
\det\,P_{j-1}(\lambda_i)}
(recall that arbitrary polynomials may be built up by adding
linear combinations of preceding columns, a procedure that leaves the
determinant unchanged),
the polynomials $P_n$ can be employed to solve \egpf. We substitute the
determinant $\det\,P_{j-1}(\lambda_i)=\sum(-1)^{\pi}
\prod_k P_{i_k-1}(\lambda_k)$
for each of the $\Delta(\lambda)$'s in \egpf\
(where the sum is over permutations $i_k$ and $(-1)^{\pi}$ is the parity of
the permutation).  The integrals over individual
$\lambda_i$'s factorize, and due to orthogonality the only contributions are
from terms with all $P_i(\lambda_j)$'s paired. There are $N!$ such terms so
\egpf\ reduces to
\eqn\egrt{\e{Z}
=N!\prod_{i=0}^{N-1}h_i=N!\, h_0^N\,\prod_{k=1}^{N-1}f_k^{N-k}\ ,}
where we have defined $f_k\equiv h_k/h_{k-1}$.

In the naive large $N$ limit (the planar limit), the rescaled index $k/N$
becomes a continuous variable $\xi$ that runs from 0 to 1, and $f_k/N$ becomes
a continuous function $f(\xi)$. In this limit, the partition function
(up to an irrelevant additive constant) reduces to a simple one-dimensional
integral:
\eqn\enlnl{{1\over N^2} Z={1\over N}\sum_k(1-k/N)\ln f_k
\sim\int_0^1\d \xi(1-\xi)\ln f(\xi)\ .}

To derive the functional form for $f(\xi)$, we assume for simplicity that the
potential $V(\lambda)$ in \eopoly\ is even.
Since the $P_i$'s from a complete set of basis vectors in the space of
polynomials, it is clear that $\lambda P_n(\lambda)$ must be expressible as a
linear combination of lower $P_i$'s, $\lambda P_n(\lambda)=\sum_{i=0}^{n+1}
a_i\,P_i(\lambda)$ (with $a_i=h_i\inv\int\ee{-V}\lambda P_n\,P_i$). In fact,
the orthogonal polynomials satisfy the simple recursion relation,
\eqn\elpn{\lambda P_n=P_{n+1}+r_n\,P_{n-1}\ ,}
with $r_n$ a scalar coefficient independent of $\lambda$. This is because any
term proportional to $P_n$ in the above vanishes due to the assumption that the
potential is even, $\int\ee{-V}\lambda\,P_n\,P_n=0$. Terms proportional to
$P_i$ for $i<n-1$ also vanish since $\int \ee{-V}P_n\,\lambda\,P_i=0$ (recall
$\lambda P_i$ is a polynomial of order at most ${i+1}$ so is orthogonal to
$P_n$ for $i+1<n$).

By considering the quantity $P_n\lambda P_{n-1}$ with $\lambda$
paired alternately with the preceding or succeeding polynomial, we derive
$$\int \ee{- V} \,P_n\, \lambda\, P_{n-1}=r_n\,h_{n-1}=h_n\ .$$
This shows that the ratio $f_n=h_n/h_{n-1}$ for this simple case\foot{In
other models, e.g.\ multimatrix models, $f_n=h_n/h_{n-1}$ has a more
complicated dependence on recursion coefficients.} is identically the
coefficient defined by \elpn, $f_n=r_n$.
Similarly if we pair the $\lambda$ in $P_n'\,\lambda\,P_n$ before and
afterwards, integration by parts gives
\eqn\enhn{n h_n=\int \ee{-V}\,P_n'\,\lambda\,  P_n
= \int \ee{-V}\,P_n'\, r_n\, P_{n-1}=r_n\int\ee{-V}\, V'\,P_n\,P_{n-1}\ .}
This is the key relation that will allow us to determine $r_n$.

\subsec{The genus zero partition function}

Our intent now is to find an expression for $f_n=r_n$ and substitute into
\enlnl\ to calculate a partition function.
For definiteness, we take as example the potential
\eqn\epex{\eqalign{&V(\lambda)={1\over 2g}\Bigl(\lambda^2+{\lambda^4\over N}
+b{\lambda^6\over N^2}\Bigr)\ ,\cr
\llap{\rm with\ derivative\quad\qquad}
g &V'(\lambda)
=\lambda+2{\lambda^3\over N}+3b{\lambda^5\over N^2}\ .\cr}}
The right hand side of \enhn\ involves terms of the form
$\int\ee{-V}\, \lambda^{2p-1}\,P_n\,P_{n-1}$. According to \elpn, these may be
visualized as ``walks'' of $2p-1$ steps ($p-1$ steps up and $p$ steps down)
starting at $n$ and ending at $n-1$, where each step down from $m$ to $m-1$
receives a factor of $r_m$ and each step up receives a factor of unity. The
total number of such walks is given by ${2p-1\choose p}$, and each results in a
 final factor of $h_{n-1}$ (from the integral
$\int\ee{-V}\,\,P_{n-1}\,P_{n-1}$) which combines with the $r_n$ to cancel the
$h_n$ on the left hand side of \enhn.
For the potential \epex, \enhn\ thus gives
\eqn\egnpx{g n=r_n+{2\over N}r_n(r_{n+1}+r_n+r_{n-1})+
{3 b\over N^2}(10 \ rrr \ {\rm terms})\ .}
(The 10 $rrr$ terms start with $r_n(r_n^2 +r_{n+1}^2+r_{n-1}^2+\ldots)$
and may be found e.g.\ in \rIYL.)

As mentioned before \enlnl, in the large $N$ limit the index $n$ becomes a
continuous variable $\xi$, and we have
$r_n/N\to r(\xi)$ and $r_{n\pm1}/N\to r(\xi\pm\varepsilon)$,
where $\varepsilon\equiv 1/N$. To leading order in $1/N$, \egnpx\ reduces to
\eqn\egxw{\eqalign{g \xi=r + 6 r^2+30 b r^3&=W(r)\cr
&=\gc+\half W''|_{r=r_c}\bigl(r(\xi)-r_c\bigr)^2+\ldots\ .\cr}}
In the second line, we have expanded $W(r)$ for $r$ near a critical point
$r_c$ at which $W'|_{r=r_c}=0$
(which always exists without any fine tuning of the parameter $b$),
and $\gc\equiv W(r_c)$. We see from \egxw\ that
$$r-r_c\sim(\gc-g \xi)^{1/2}\ .$$
(For a general potential $V(\lambda)={1\over2g}\sum_p a_p\,\lambda^{2p}$ in
\epex, we would have $W(r)=\sum_p a_p{(2p-1)!\over (p-1)!^2}\,r^p$.)

To make contact with the 2d gravity ideas of the preceding section, let us
suppose more generally that the leading singular behavior of $f(\xi)$
$\bigl(=r(\xi)\bigr)$ for large $N$ is
\eqn\efx{f(\xi)-f_c \sim (\gc - g \xi)^{-\gamma}}
for $g$ near some $\gc$ (and $\xi$ near 1). (We shall see that $\gamma$ in the
above coincides with the critical exponent $\gamma$ defined in \elpoa.)
The behavior of \enlnl\ for $g$ near $\gc$ is then
\eqn\emc{\lbspace\eqalign{{1\over N^2}Z
\sim\int_0^1\d \xi\,(1-\xi)(\gc- g \xi)^{-\gamma}
&\sim(1-\xi)(\gc-g \xi)^{-\gamma+1}\Big|_0^1
+\int_0^1\d \xi\,(\gc- g \xi)^{-\gamma+1}\cr
&\sim(\gc-g)^{-\gamma+2}\sim \sum_n n^{\gamma-3}(g/\gc)^n\ .}}
Comparison with \elpoa\ shows that the large area (large $n$) behavior
identifies the exponent $\gamma$ in \efx\ with the critical exponent defined
earlier. We also note that the second derivative of $Z$ with respect to
$x=\gc-g$ has leading singular behavior
\eqn\elsbz{Z''\sim(\gc-g)^{-\gamma}\sim f(1)\ .}

{}From \efx\ and \emc\ we see that the behavior in \egxw\ implies a critical
exponent $\gamma=-1/2$. From \epbl, we see that this corresponds to the case
$D=0$, i.e.\ to pure gravity. It is natural that pure gravity should be present
for a generic potential. With fine tuning of the parameter
$b$ in \epex, we can achieve a higher order critical point, with
$W'|_{r=r_c}=W''|_{r=r_c}=0$, and hence the r.h.s.\ of \egxw\ would instead
begin with an $(r-r_c)^3$ term. By the same argument starting from \efx, this
would result in a critical exponent $\gamma=-1/3$. With a general potential
$V(M)$ in \egpf, we have enough parameters to achieve an $m^{\rm th}$ order
critical point \rkazcon\ at which the first $m-1$ derivatives of $W(r)$ vanish
at $r=r_c$. The behavior is then
$r-r_c\sim (\gc-g \xi)^{1/m}$ with associated critical exponent $\gamma=-1/m$.
As anticipated at the end of subsection {\it 1.2\/}, we see that more general
polynomial matrix interactions provide the necessary degrees of freedom to
result in matter coupled to 2d gravity in the continuum limit.

\subsec{The all genus partition function}

We now search for another solution to \egnpx\ and its generalizations that
describes the contribution of all genus surfaces to the partition function
\enlnl. We shall retain higher order terms in $1/N$ in \egnpx\ so that e.g.\
\egxw\ instead reads
\eqn\egxwp{\eqalign{g \xi
&=W(r)+ 2 r(\xi)\bigl(r(\xi+\varepsilon)+r(\xi-\varepsilon)-2r(\xi)\bigr)\cr
&=\gc+\half W''|_{r=r_c}\bigl(r(\xi)-r_c\bigr)^2
+ 2 r(\xi)\bigl(r(\xi+\varepsilon)+r(\xi-\varepsilon)-2r(\xi)\bigr)+\ldots\
.\cr}}
As suggested at the end of section 1, we shall simultaneously let $N\to\infty$
and $g\to\gc$ in a particular way. Since $g-\gc$ has dimension [length]$^2$,
it is convenient to introduce a parameter $a$ with dimension length and
let $g-\gc=\kappa^{-4/5}a^2$, with $a\to0$. Our ansatz for a coherent
large $N$ limit will be to take $\varepsilon\equiv 1/N=a^{5/2}$
so that the quantity $\kappa\inv=(g-\gc)^{5/4}N$ remains finite as $g\to\gc$
and $N\to\infty$.

Moreover since the integral \enlnl\ is dominated by
$\xi$ near 1 in this limit, it is convenient to change variables from $\xi$ to
$z$, defined by $\gc-g\xi=a^2 z$.
Our scaling ansatz in this region is $r(\xi)=r_c+a u(z)$.
If we substitute these definitions into \egxw, the leading terms are of order
$a^2$ and result in the relation $u^2\sim z$. To include the higher derivative
terms, we note that
$$r(\xi+\varepsilon)+r(\xi-\varepsilon)-2r(\xi)
\sim \varepsilon^2{\del^2 r\over \del \xi^2}
=a {\del^2\over \del z^2} a u(z)\sim a^2 u''\ ,$$
where we have used
$\varepsilon(\del/\del \xi)=-g a^{1/2}(\del/\del z)$
(which follows from the above change of variables from $\xi$ to $z$).
Substituting into \egxwp, the vanishing of the coefficient of $a^2$ implies
the differential equation
\eqn\eplv{z=u^2-{\textstyle 1\over3}u''}
(after a suitable rescaling of $u$ and $z$).
In \elsbz, we saw that the
second derivative of the partition function
(the ``specific heat'') has leading singular behavior given by
$f(\xi)$ with $\xi=1$, and thus by $u(z)$ for $z=(g-\gc)/a^2=\kappa^{-4/5}$.
The solution to \eplv\ characterizes the behavior of the partition function of
pure gravity to all orders in the genus expansion. (Notice that the leading
term is $u\sim z^{1/2}$ so after two integrations the leading term in $Z$ is
$z^{5/2}=\kappa^{-2}$, consistent with \elner.)

Eq.~\eplv\ is known in the mathematical literature as the
Painlev\'e I equation.
The perturbative solution in powers of $z^{-5/2}=\kappa^2$ takes the form
$u=z^{1/2}(1-\sum_{k=1}u_k z^{-5k/2})$,
where the $u_k$ are all positive.\foot{The first term, i.e.\ the
contribution from the sphere, is dominated by a regular part which has
opposite sign.  This is removed by taking an additional derivative of $u$,
giving a series all of whose terms have the same sign ---
negative in the conventions of \eplv. The other solution, with leading term
$-z^{1/2}$, has an expansion with alternating sign
which is presumably Borel summable, but not physically relevant.}
This verifies for this model the claims made in eqs.~\elolnhg--\elner\ of
subsection {\it 1.4\/}.
For large $k$, the $u_k$ go asymptotically as $(2k)!$, so the solution for
$u(z)$ is not Borel summable (for a review of these issues in the context of 2d
gravity, see e.g.\ \rGZaplob). Our arguments in section 1 show only that the
matrix model results should agree with 2d gravity order by order in
perturbation theory. How to insure that we are studying nonperturbative gravity
as opposed to nonperturbative matrix models is still an open question.
Some of the constraints that the solution to \eplv\ should satisfy are
reviewed in \rfrdc. In particular it is known that real solutions to \eplv\
cannot satisfy the Schwinger-Dyson (loop) equations for the theory.

In the case of the next higher multicritical point, with $b$ in \egxw\ adjusted
so that $W'=W''=0$ at $r=r_c$, we have
$W(r)\sim \gc+{1\over6}W'''|_{r=r_c}(r-r_c)^3+\ldots$
and critical exponent $\gamma=-1/3$.
In general, we take $g-\gc=\kappa^{2/(\gamma-2)}a^2$, and
$\varepsilon=1/N=a^{2-\gamma}$ so that the combination
$(g-\gc)^{1-\gamma/2}N=\kappa\inv$ is fixed in the limit $a\to0$.
The value  $\xi=1$ now corresponds to $z=\kappa^{2/(\gamma-2)}$, so the
string coupling $\kappa^2=z^{\gamma-2}$.
The general scaling scaling ansatz is
$r(\xi)=r_c+a^{-2 \gamma}u(z)$, and the change of variables from $\xi$ to $z$
gives $\varepsilon(\del/\del \xi)=-g a^{-\gamma}(\del/\del z)$.

For the case $\gamma=-1/3$, this means in particular that
$r(\xi)=r_c+a^{2/3}u(z)$, $\kappa^2=z^{-7/3}$, and
$\varepsilon(\del/\del \xi)=-ga^{1/3}{\del\over \del z}$.
Substituting into the large $N$ limit of \egnpx\ gives
(again after suitable rescaling of $u$ and $z$)
\eqn\etmcp{z=u^3-u u'' -\half(u')^2+\alpha\, u''''\ ,}
with $\alpha={1\over10}$. The solution to \etmcp\ takes the form
$u=z^{1/3}(1+\sum_k u_k\,z^{-7k/3})$. It turns out that the coefficients $u_k$
in the perturbative expansion of the solution to \etmcp\ are positive definite
only for $\alpha<{1\over12}$, so the $3^{\rm th}$
order multicritical point does not
describe a unitary theory of matter coupled to gravity. Although
from \epbl\ we see that
the critical exponent $\gamma=-1/3$ coincides with that predicted for the
(unitary) Ising model coupled to gravity,
it turns out \refs{\rIYL,\rising}\ that \etmcp\ with $\alpha={1\over10}$
instead describes the conformal field theory of the Yang-Lee edge singularity
(a critical point obtained by coupling the Ising model to a particular value of
imaginary magnetic field) coupled to gravity. The specific heat of the
conventional critical Ising model coupled to gravity turns out (see the next
section here) to be as well determined by the differential equation \etmcp, but
instead with $\alpha={2\over27}$.

For the general $m^{\rm th}$ order critical point of the potential $W(r)$, we
have seen that the associated model of matter coupled to gravity has critical
exponent $\gamma=-1/m$.
With scaling ansatz $r(\xi)=r_c+a^{2/m}u(z)$, we find leading behavior
$u\sim z^{1/m}$ (and $Z\sim z^{2+1/m}=\kappa^{-2}$ as expected).
The differential equation that results from substituting the double scaling
behaviors given before \etmcp\ into the generalized version of \egnpx\ turns
out to be the $m^{\rm th}$ member of the KdV hierarchy of differential
equations (of which Painlev\'e I results for $m=2$). In the next section, we
shall provide some marginal insight into why this structure emerges.

In the nomenclature of \rBPZ, so-called ``minimal conformal field theories''
(those with a finite number of primary fields) are specified by a pair of
relatively prime integers $(p,q)$. (The unitary discrete series is the subset
specified by $(p,q)=(m+1,m)$.)
After coupling to gravity, these have critical exponent $\gamma=-2/(p+q-1)$.
In general, the $m^{\rm th}$ order multicritical point of the one-matrix model
turns out to describe the $(2m-1,2)$ model (in general non-unitary) coupled to
gravity, so its critical exponent $\gamma=-1/m$ happens to coincide with that
of the $m^{\rm th}$ member of the unitary discrete series coupled to gravity.
The remaining $(p,q)$ models coupled to gravity can be realized in terms of
multi-matrix models (to be defined in the next section).

\newsec{KdV equations and other models}

\subsec{KdV equations}

We now wish to describe superficially why the KdV hierarchy of differential
equations plays a role in 2d gravity. To this end it is convenient to switch
from the basis of orthogonal polynomials $P_n$ employed in the previous section
to a basis of orthonormal polynomials
$\Pi_n(\lambda) = P_n(\lambda)/\sqrt{h_n}$ that satisfy
\eqn\eonP{\int_{-\infty}^\infty\d \lambda\, \ee{-V}\,\Pi_n\,\Pi_m
=\delta_{nm}\ .}
In terms of the $\Pi_n$, eq.~\elpn\ becomes
$$\eqalign{\lambda\Pi_n&=\sqrt{h_{n+1}\over h_n}\,\Pi_n
+r_n\sqrt{h_{n-1}\over h_n}\,\Pi_{n-1}
=\sqrt{r_{n+1}}\,\Pi_{n+1}+\sqrt{r_n}\,\Pi_{n-1}\cr
&=Q_{nm}\,\Pi_m\ .\cr}$$
In matrix notation, we write this as $\lambda\Pi= Q\Pi$,
where the matrix $Q$ has components
\eqn\eQmn{Q_{nm}=\sqrt{r_m}\delta_{m,n+1}+\sqrt{r_n}\delta_{m+1,n}\ .}
Due to the orthonormality property \eonP, we see that $\int\ee{-V}\lambda
\Pi_n\,\Pi_m=Q_{nm}=Q_{mn}$, and
$Q$ is a symmetric matrix. In the continuum limit, $Q$ will therefore become
a hermitian operator.

To see how this works explicitly \refs{\rD,\rbdss},
we substitute the scaling ansatz
$r(\xi)=r_c+a^{2/m}u(z)$ for the $m^{\rm th}$ multicritical model into \eQmn,
$$Q\to(r_c+a^{2/m}u(z))^{1/2}\,\e{\varepsilon{\del\over\del \xi}}
+\e{-\varepsilon{\del\over\del \xi}}(r_c+a^{2/m}u(z))^{1/2}\ .$$
With the substitution
$\varepsilon{\del\over\del \xi}\to -g a^{1/m}{\del\over\del z}$, we find the
leading terms
\eqn\eQmnc{Q=2 r_c^{1/2}+{a^{2/m}\over\sqrt{r_c}}(u+r_c \kappa^2\del_z^2)\ ,}
of which the first is a non-universal constant and the second is a hermitian
$2^{\rm nd}$ order differential operator.

The other matrix that naturally arises is defined by differentiation,
\eqn\edA{{\del\over\del \lambda}\Pi_n=A_{nm}\Pi_m \ ,}
and automatically satisfies $[A,Q]=1$. The matrix $A$ does not have any
particular symmetry or antisymmetry properties so it is convenient to correct
it to a matrix $P$ that satisfies the same commutator as $A$.
{}From our definitions, it follows that
$$0=\int{\del\over\del \lambda}\bigl(\Pi_n\,\Pi_m\,\ee{-V}\bigr)\quad
\quad\Rightarrow\quad A+A^T=V'(Q)\ ,$$
where we have differentiated term by term and used
$\int\ee{-V}\lambda^\ell\, \Pi_n\,\Pi_m=(Q^\ell)_{nm}$.
The matrix $P\equiv A-\ha V'(Q)=\ha(A-A^T)$ is therefore anti-symmetric and
satisfies
\eqn\epqd{\bigl[P,Q\bigr]=1\ .}

To determine the order of the differential operator $Q$ in the continuum
limit, let us assume for example that the potential $V$ is of order
$2\ell$, i.e.\ $V=\sum_{k=0}^{\ell}a_k\,\lambda^{2k}$. For $m>n$, the
integral $A_{mn}=\int \ee{-V}\Pi_n{\del\over\del \lambda}\Pi_m=\int
\ee{-V} V'\,\Pi_n\,\Pi_m$ may be nonvanishing for $m-n\le 2\ell-1$. That
means that $P_{mn}\ne0$ for $|m-n|\le 2\ell-1$, and thus has enough
parameters to result in a $(2\ell-1)^{\rm st}$ order differential operator
in the continuum. The single condition $W'=0$ results in $P$ tuned to a
$3^{\rm rd}$ order operator, and the $\ell-1$ conditions
$W'=\ldots=W^{(\ell-1)}=0$ allow $P$ to be realized as a $(2\ell-1)^{\rm st}$
order differential operator. In \eQmnc, we see that the universal
part of $Q$ after suitable rescaling takes the form $Q=\d^2-u$. For the
simple critical point $W'=0$, the continuum limit of $P$ is the
antihermitian operator $P=\d^3-{3\over4}\{u,\d\}$, and the commutator
\eqn\ert{1=[P,Q]=4R_2'=
\Bigl({\textstyle 3\over4}u^2-{\textstyle 1\over4}u''\Bigr)'}
is easily integrated with respect to $z$ to give an equation equivalent to
\eplv, the string equation for pure gravity. In \ert, the notation $R_2$ is
conventional for the first member of the ordinary KdV hierarchy. The emergence
of the KdV hierarchy in this context is due to the natural occurrence of the
fundamental commutator relation \epqd, which also occurs in the Lax
representation of the KdV equations. (The topological gravity approach has as
well been shown at length to be equivalent to KdV, for a review see \rdvv.)

In general the differential equations
\eqn\epqcc{[P,Q]=1}
that follow from \epqd\
may be determined directly in the continuum. Given an operator $Q$,
the differential operator $P$ that can satisfy this commutator is constructed
as a ``fractional power'' of the operator $Q$.

Before showing how this construction works, we first expand slightly the class
of models from single matrix to multi-matrix models.
The free energy of a particular $(q-1)$-matrix model, generalizing \egpf, may
be written \rCMM\
\eqn\eqmm{\eqalign{Z&=\ln\int\prod_{i=1}^{q-1} \d M_i\
\e{-\tr\bigl(\sum_{i=1}^{q-1}V_i(M_i)
-\sum_{i=1}^{q-2} c_i\,M_i M_{i+1}\bigr)}\cr
&=\ln\int \prod_{{\scriptstyle i=1,q-1}\atop {\scriptstyle\alpha=1,N}}
\!\!\d\lambda_i^{(\alpha)}\ \Delta(\lambda_1)\,
\e{-\sum_{i,\alpha}V_i\bigl(\lambda_i^{(\alpha)}\bigr)+
\sum_{i,\alpha} c_i\,\lambda_i^{(\alpha)}\lambda_{i+1}^{(\alpha)}}
\Delta(\lambda_{q-1})\ ,\cr}}
where the $M_i$ (for $i=1,\ldots,q-1$) are $N\times N$ hermitian matrices,
the $\lambda_i^{(\alpha)}$ ($\alpha=1,\ldots,N$) are their eigenvalues, and
$\Delta(\lambda_i)=\prod_{\alpha<\beta}(\lambda_i^{(\alpha)}-
\lambda_i^{(\beta)})$ is the Vandermonde determinant.
The result
in the second line of \eqmm\ depends on having $c_i$'s that couple matrices
along a line (with no closed loops so that the integrations over the relative
angular variables in the $M_i$'s can be performed.)
Via a diagrammatic expansion, the matrix integrals in \eqmm\
can be interpreted to generate a sum over discretized surfaces, where
the different matrices $M_i$ represent $q-1$ different matter states that
can exist at the vertices.
The quantity $Z$ in \eqmm\ thereby admits an interpretation as the partition
function of 2d gravity coupled to matter.

Following \rCMM, we can
introduce operators $Q_i$ and $P_i$ that represent the insertions of
$\lambda_i$ and $\d/\d\lambda_i$ respectively in the integral \eqmm.
These operators necessarily satisfy $\bigl[P_i,Q_i\bigr]=1$.
In the $N\to\infty$ limit, we have seen (following \rD) that $P$ and $Q$
become differential operators of finite order, say $p,q$ respectively
(where we assume $p>q$), and these continue to satisfy \epqcc.
\def\QT{K}
In the continuum limit of the matrix problem (i.e.\ the ``double'' scaling
limit, which here means couplings in \eqmm\ tuned to
critical values), $Q$ becomes a differential operator of the form
\eqn\eQ{Q=\d^{q}+\left\{v_{q-2}(z),\d^{q-2} \right\}+\ \cdots\ + 2v_{0}(z)\ ,}
where $\d=\d/\d z$. (By a change of basis of the form $Q\to f\inv(z)Qf(z)$,
the coefficient of $\d^{q-1}$ may always be set to zero.)
The continuum scaling limit of the multi-matrix models
is thus abstracted to the mathematical problem of finding solutions to \epqcc.

The differential equations \epqcc\ may be constructed as follows.
For $p,q$ relatively prime, a $p^{\rm th}$ order
differential operator that can satisfy \epqcc\ is
constructed as a fractional power of the operator $Q$ of \eQ.
Formally, a $q^{\rm th}$ root may be represented within an algebra of
formal pseudo-differential operators (see, e.g.\
\ref\rDrSo{V. G. Drinfel'd and V. V. Sokolov, Jour. Sov. Math. (1985) 1975\semi
G. Segal and G. Wilson, Pub. Math. I.H.E.S. 61 (1985), 5.}) as
\eqn\eQr{Q^{1/q}=\d + \sum_{i=1}^{\infty} \left\{e_i,\d^{-i}\right\}\ ,}
where $\d\inv$ is defined to satisfy
$\d\inv f=\sum_{j=0}^\infty (-1)^j f^{(j)}\,\d^{-j-1}$.
The differential equations describing the $(p,q)$ minimal model coupled to 2d
gravity are given by
\eqn\ecom{\bigl[Q^{p/q}_+,\ Q\bigr]=1\ ,}
where $P=Q^{p/q}_+$ indicates the part of $Q^{p/q}$ with only
non-negative powers of d, and is a $p^{\rm th}$ order differential operator.

To illustrate the procedure we reproduce now the results for the one-matrix
models, which can be used to generate $(p,q)$ of the form $(2l-1,2)$.
{}From \eQmnc, these models are obtained by taking $Q$ to be the hermitian
operator
\eqn\eQom{Q=\QT\equiv\d^2 - u(z)\ .}
The formal expansion of $Q^{l-1/2}=\QT^{l-1/2}$
(an anti-hermitian operator) in powers of $\d$ is given by
\eqn\eDl{\QT^{l-1/2}=\d^{2l-1}- {2l-1\over4}\left\{u,\d^{2l-3}\right\}
+ \ldots\ }
(where only symmetrized odd powers of $\d$ appear in this case).
We now decompose $\QT^{l-1/2}= \QT^{l-1/2}_+ + \QT^{l-1/2}_-$, where
$\QT^{l-1/2}_+=\d^{2l-1}+\ldots$ contains only non-negative powers of $\d$,
and the remainder $\QT^{l-1/2}_-$ has the expansion
\eqn\eQR{\QT^{l-1/2}_-=\sum_{i=1}^\infty\
\bigl\{e_{2i-1},\d^{-(2i-1)}\bigr\}
=\left\{R_{l},\d^{-1}\right\}+O(\d^{-3})+\ldots\ .}
Here we have identified $R_l\equiv e_1$ as the first term in the expansion
of $\QT^{l-1/2}_-$.
For $\QT^{1/2}$, for example, we find $\QT^{1/2}_+=\d$ and $R_1=-u/4$.

The prescription \ecom\ with $p=2l-1$ corresponds here to
calculating the commutator $\bigl[\QT^{l-1/2}_+,\QT\bigr]$.
Since $\QT$ commutes with $\QT^{l-1/2}$, we have
\eqn\eKKc{\bigl[\QT^{l-1/2}_+,\QT\bigr]=\bigl[\QT,\QT^{l-1/2}_-\bigr]\ .}
But since $\QT$ begins at $\d^2$, and since from the l.h.s.\ above
the commutator can have only positive powers of $\d$,
only the leading ($\d\inv$) term from the r.h.s.\ can contribute, which
results in
\eqn\elc{\bigl[\QT^{l-1/2}_+,\QT\bigr]
=\,{\rm leading\ piece\ of}\ \bigl[\QT,2 R_{l}\,\d^{-1}\bigr]
=4R'_{l}\ .}
After integration, the equation $\bigl[\QT^{l-1/2}_+,\QT\bigr]=1$
thus takes the simple form
\eqn\ede{c\,R_{l}[u]=z\ ,}
where the constant $c$ may be fixed by suitable rescaling of $z$ and $u$
(enabled by the property that all terms in $R_l$ have fixed grade, namely
$2l$).

The quantities $R_l$ in \eQR\ are easily seen to satisfy a simple recursion
relation. From $\QT^{l+1/2}=\QT\QT^{l-1/2}=\QT^{l-1/2}\QT$, we find
%
$$\QT^{l+1/2}_+=\ha\left(\QT^{l-1/2}_+ \QT
+ \QT\QT^{l-1/2}_+\right)+\bigl\{R_l,\d\bigr\} \ .$$
%
Commuting both sides with $\QT$ and using \elc, simple algebra gives \rGD
\eqn\erec{R'_{l+1}={1\over4}R'''_{l}-uR'_{l}-{1 \over 2}u'R_l\ .}

While this recursion formula only determines $R'_{l}$,
by demanding that the $R_{l}$ ($l\ne0$) vanish at $u=0$, we obtain
\eqn\erex{\eqalign{R_0&=\ha\,,\qquad\qquad
R_1=-{1\over4}u\,,\qquad\qquad
R_2={3\over16}u^2-{1\over16}u''\,,\cr
R_3&=-{5\over32}u^3+{5\over32}\bigl(uu''+\half u'{}^2\bigr)
-{1\over64}u^{(4)}\ .\cr}}
We summarize as well the first few $\QT^{l-1/2}_+$,
\eqn\ekex{\eqalign{\QT^{1/2}_+&=\d\,,\qquad\qquad
\QT^{3/2}_+=\d^3-{3\over4}\{u,\d\}\,,\cr
\QT^{5/2}_+&=\d^5-{5\over4}\{u,\d^3\}+
{5\over16}\left\{(3u^2+u''),\d \right\}\ .\cr}}

After rescaling, we recognize $R_3$ in \erex\ as eq.~\etmcp\
with $\alpha={1\over10}$, i.e. the equation for the (2,5) model. In general,
the equations determined by \epqcc\ for general $p,q$ characterize the
partition function of the $(p,q)$ minimal model (mentioned at the end of
section 2) coupled to gravity. To realize these equations in the continuum
limit turns out \refs{\rmrdt,\rtadt} to require only a two-matrix model of the
type \eqmm. The argument given after \epqd\ for the one-matrix case is easily
generalized to the recursion relations for the two-matrix case and shows that
for high enough order potentials, there are enough couplings to tune
the matrices $P$ and $Q$ to become $p^{\rm th}$ and $q^{\rm th}$ order
differential operators. In subsection {\it 3.2\/}, we shall show how to realize
a $D=1$ theory coupled to gravity in terms of a two-matrix model.
In \rtmr, it is argued that one can as well realize a wide variety of
$D<1$ theories by means of a one-matrix model coupled to an external potential.

\subsec{Other models}

As a specific example of a two-matrix model, we consider
\eqn\eIs{\e{Z}=\int \d U\,\d V\
\e{-\tr\bigl(U^2+V^2-2c\, UV+{g\over N}(\ee{H}\,U^4+\ee{-H}V^4)\bigr)}\ ,}
where $U$ and $V$ are hermitian $N\times N$ matrices and $H$ is a constant.
In the diagrammatic expansion of the right hand side, we now have two different
quartic vertices of the type depicted in fig.~3b, corresponding to insertions
of $U^4$ and $V^4$. The propagator is determined by the inverse of the
quadratic term,
$$\pmatrix{1&-c\cr -c&1\cr}\inv={1\over 1-c^2}\pmatrix{1&c\cr c&1\cr}\ .$$
We see that double lines connecting vertices of the same type (either generated
by $U^4$ or $V^4$) receive a factor of $1/(1-c^2)$, while those connecting
$U^4$ vertices to $V^4$ vertices receive a factor of $c/(1-c^2)$.

This is identically the structure necessary to realize the Ising model on a
random lattice. Recall that the Ising model is defined to have a spin
$\sigma=\pm1$ at each site of a lattice, with an interaction $\sigma_i
\sigma_j$ between nearest neighbor sites $\langle ij\rangle$. This interaction
takes one value for equal spins and another value for unequal spins. Up to an
overall additive constant to the free energy, the diagrammatic expansion of
\eIs\ results in the 2d partition function
$$Z=\sum_{\rm lattices}
\sum_{{\scriptstyle\rm spin}\atop{\scriptstyle\rm configurations}}
\e{\beta\sum_{\langle ij\rangle}\sigma_i\,\sigma_j+H\sum_i \sigma_i}$$
where $H$ is the magnetic field. The weights for equal and unequal neighboring
spins are $\ee{\pm \beta}$, so fixing the ratio $\ee{2 \beta}=1/c$ relates the
parameter $c$ in \eIs\ to the temperature $\beta$. It turns out that the Ising
model is much easier to solve summed over random lattices than on a regular
lattice, and in particular is solvable even in the presence of a magnetic
field. This is because there is much more symmetry after coupling to gravity,
since the complicating details of any particular lattice (e.g.\ square) are
effectively integrated out.

We briefly outline the method for solving \eIs\ (see \refs{\rKBK,\rIYL,\rising}
for more details). By methods similar to those used to derive \egpf, we can
write \eIs\ in terms of the eigenvalues $x_i$ and $y_i$ of $U$ and $V$,
$$\e{Z}=\int \Delta(x)\,\Delta(y)\ \prod_i\d x_i\,\d y_i\,\e{-W(x_i, y_i)}
\ .$$
where $W(x_i, y_i)\equiv x_i^2+y_i^2-2c\,x_iy_i
+{g\over N}(\ee{H}x_i^4+\ee{-H}y_i^4)$.
The polynomials we define for this problem are orthogonal with respect to the
bilocal measure
$$\int \d x\,\d y\ \ee{-W(x,y)}\,P_n(x)\,Q_m(y)=h_n\,\delta_{nm}$$
(where $P_n\ne Q_n$ for $H\ne0$). The result for the partition function is
identical to \egrt,
$$\e{Z}\propto\prod_i h_i\propto\prod_i f_i^{N-i}\ ,$$
and the recursion relations for this case generalize \elpn,
$$\eqalign{x\,P_n(x)&=P_{n+1}+r_n\,P_{n-1}+s_n\,P_{n-3}\ ,\cr
y\,Q_m(y)&=Q_{m+1}+q_m\,Q_{m-1}+t_m\,Q_{m-3}\ .\cr}$$
We still have $f_n\equiv h_n/h_{n-1}$, and $f_n$ can be determined in terms of
the above recursion coefficients (although the formulae are more complicated
than in the one-matrix case). After we substitute
the scaling ans\"atze described in subsection {\it 2.3\/}, the formula for the
scaling part of $f$ is derived via straightforward algebra. The result
is that the specific heat $u\propto Z''$ is given by \etmcp\ with
$\alpha={2\over27}$.

Other conventional statistical mechanical models can be formulated on
random lattices and solved in the continuum limit. The ADE face models (with
$D<1$), for example, have been considered in \rkade.
One way of formulating $D=1$ is to generalize \eqmm\ to an infinite line of
matrices. In dual form, this is equivalent to strings propagating on a circle
of finite radius (see e.g.\ \refs{\rdOne,\rdOneGK}). Another formulation
involves letting the index $i$ specifying the matrix $M_i$ become a continuous
index $t\in(-\infty,\infty)$. In this limit we trade off matrix quantum
mechanics for a field theory of matrices theory $M(t)$. This is a problem that
was originally solved in \rBIPZ, and was used to analyze 2d gravity at genus
zero in \rKM\ and was then applied to higher genus starting in
\refs{\rdOne,\rGZ}. A connection to Liouville theory was pointed out in
\rPcbrsod, and carried further by the
free fermion and collective field formulations of \rColl.

Yet another means of formulating 2d gravity coupled to $D=1$ matter is via the
8-vertex model, which renormalizes at criticality (the 6-vertex model)
onto a single boson at finite radius.\foot{For an overview geared towards
string/particle physicists, see e.g.\ \rpgtr. On a regular lattice, the
radius $r$ of the boson (in conventions in which $r=1/\sqrt2$ is the
self-dual point) and the conventional weights $a,b,c$ of the 6-vertex
model are related by $\cos\ha\pi r^2=(a^2+b^2-c^2)/2ab$.} Since this has
not been treated in the literature, we give a quick description of the
formulation. The simplest vertex models are those for which the degrees of
freedom are (two-state) arrows that live on links, and are defined on
lattices which have four links meeting at each vertex. Each possible arrow
configuration at a vertex is given a statistical weight, and the partition
function is given by summing over all arrow configurations, with each
assigned an overall weight equal to the product of the statistical weights
over the vertices. In the 8-vertex model, the vertices are restricted to
the set of eight with an even number of arrows both incoming and outgoing.
In the 6-vertex model, the source and sink (all four arrows outgoing or
incoming) are excluded, which leaves the four distinct rotated versions of
fig.~4a, and the two distinct rotated versions of fig.~4b.

$$\hbox{\vbox{\hbox{$\textfont2\bigarrfont\mathord\rightarrow\mkern-6mu\extarr
\vuline\hskip-4.75pt\lower23pt\hbox{\vuline}
\mkern-1mu\mathord\rightarrow\mkern-6mu\extarr$}
\vskip7pt\hbox{$\!\!$\qquad(a)}}
\qquad\qquad\qquad
\vbox{\hbox{$\textfont2\bigarrfont\mathord\rightarrow\mkern-6mu\extarr
\vuline\hskip-4.75pt\lower23pt\hbox{\vdline}
\mkern-2mu\extarr\mathord\leftarrow\mkern-6mu$}
\vskip7pt\hbox{$\!\!$\qquad(b)}}}$$
\nobreak
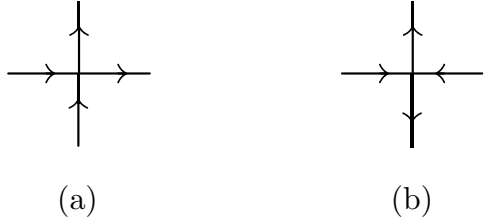
\centerline{\footnotefont{\bf Fig.~4:} (a) vertex with weight $a$. \quad
(b) vertex with weight $c$.}
\bigbreak

The coupling to gravity is given by summing over random lattices that maintain
four links at each vertex, but can have arbitrary polygonal faces. It is simple
to write down a matrix model that generates 6-vertex configurations on random
lattices. Rather than a hermitian matrix, we employ an arbitrary complex
$N\times N$ matrix $\ph=A+iB$,
where $A$ and $B$ are hermitian. The propagator $\langle\ph^\dagger\ph\rangle$
now has an overall orientation, which we identify by an arrow on the
propagator.
(In what follows we suppress the underlying double-lined notation of fig.~3.)
The graphs of interest are generated by the matrix integral
\eqn\esvm{\int_\ph \e{\tr\bigl(-\ha\ph^\dagger\ph+a\ph^2\ph^{\dagger 2}
+c(\ph^\dagger\ph)^2\bigr)}\ ,}
where the vertices shown in figs.~4a and 4b are assigned weights $a$ and $c$
respectively.\foot{On a regular square lattice, the four rotated versions of
fig.~4a are further subdivided into two mirror reflected pairs, which can be
given different weights $a$ and $b$. On a random lattice such distinctions are
academic, a property embodied by the cyclicity of the trace in \esvm, and we
automatically generate the so-called F-model with $a=b$.}

The model has not yet been solved in this formulation except at the analog of
the Kosterlitz-Thouless point, $a=c$. At that point we can use the identity
$$\tr\bigl[\ph^2\ph^{\dagger 2}+(\ph^\dagger\,\ph)^2\bigr]
={1\over8}\bigl[(\ph+\ph^\dagger)^2-(\ph-\ph^\dagger)^2\bigr]^2
=2\,\tr(A^2+B^2)^2$$
to rewrite the action in terms of the hermitian matrices $A,B$. By introducing
an additional integration matrix $M$, we can reduce the action to terms
quadratic in $A$ and $B$,
$$\e{2c\,\tr\bigl(A^2+B^2\bigr)^2}
=\int_M\e{\tr\bigl(-\ha M^2+2\sqrt c(A^2+B^2)M\bigr)}\ .$$
In this form, the model reduces to a standard transcription of the $O(n)$ model
for $n=2$. (For general $O(n)$, $A^2+B^2$ is replaced in the above by
$\sum_{i=1}^n A_i^2$.) This is reasonable since $SO(2)$ is just the circle
$S^1$ normalized to a particular radius. The genus zero solution (due to M.
Gaudin) is reproduced in \rkco.

\newsec{Quick tour of Liouville theory}

For completeness, we give here a brief overview\foot{In preparing this
section, I may have shamelessly plagiarized some material from a similar
section in \rAGnotes\ (whose author is consequently responsible for any
conceptual errors contained herein). Historically, after the work of
\rpoly\ some of the results here where derived in \rctbg, where the
conformal quantization of Liouville theory was studied (but general
correlation functions were not calculated). The quantum Liouville theory
was also studied in \rGN. More recently, the calculation of critical
exponents in lightcone gauge was carried out in \rKPZ\ (using $SL(2,\IR)$
current algebra). The results were subsequently rederived in conformal
gauge in \rDDK, which is the approach we follow here since it applies also
to higher genus. Reviews of Liouville theory may be found in \rSliouv.} of
how the continuum results we have used here are calculated. As previously
mentioned, the coincidence of these results with those of the matrix model
approach originally served to give post-facto verification of both
methods. This section may be considered as an appendix to the preceding
three.

\subsec{String susceptibility $\gamma$}

We consider the continuum partition function
\eqn\epf{Z=\int {\CD g\,\CD X\over {\rm Vol(Diff)}}\
\e{-S_M(X;g)-{\mu_0\over2\pi} \int \d^2\xi\,\sqrt g}\ ,}
where $S_M$ is some conformally invariant action for matter fields coupled to a
two dimensional surface $\Sigma$ with metric $g$, $\mu_0$ is a bare
cosmological constant, and we have symbolically divided the measure by the
``volume'' of the diffeomorphism group (which acts as a local symmetry) of
$\Sigma$ . For the free bosonic string, we take
$S_M={1\over8\pi}\int \d^2\xi\,\sqrt g\, g^{ab}\,
\del_a \vec X\cdot\del_b\vec X$ where the $\vec X(\xi)$ specify the embedding
of $\Sigma$ into flat $D$-dimensional spacetime.

To define \epf, we need to specify the measures for the integrations over
$X$ and $g$ (see, e.g.\ \rFrlh). The measure $\CD X$ is determined by requiring
that $\int \CD_g \delta X\ \ee{-\|\delta X\|^2_g}=1$, where the norm in the
gaussian functional integral is given by
$\|\delta X\|^2_g=\int \d^2\xi\,\sqrt g\,\delta\vec X\cdot\delta\vec X$.
Similarly, the measure $\CD g$ is determined by normalizing
$\int\CD_g \delta g\ \ee{-\half\|\delta g\|^2_g}=1$, where
$\|\delta g\|^2_g=\int \d^2\xi\,\sqrt g\,
(g^{ac}g^{bd}+2g^{ab}g^{cd})\,\delta g\dup_{ab}\,\delta g\dup_{cd}$, and
$\delta g$ represents a metric fluctuation at some point
$g\dup_{ij}$ in the space of metrics on a genus $h$ surface.

The measures $\CD X$ and $\CD g$ are invariant under the group of
diffeomorphisms of the surface, but not necessarily under conformal
transformations $g\dup_{ab}\to\ee\sigma\,g\dup_{ab}$.
Indeed due to the metric dependence in the norm $\|\delta X\|^2_g$, it turns
out that
\eqn\edx{\CD_{e^\sigma\! g}X = \e{{D\over 48\pi}\, S_L(\sigma)} \CD_g X\ ,}
where
\eqn\eL{S_L( \sigma)=\int \d^2\xi\,\sqrt g\,\Bigl(\half g^{ab}\, \del_a\sigma
\del_b\sigma + R\sigma + \mu\ee\sigma\Bigr)}
is known as the Liouville action.
(This result may be derived diagrammatically, via the Fujikawa method, or via
an index theorem; for a review see \rOA.)

The metric measure $\CD g$ as well has an anomalous variation under conformal
transformations. To express it in a form analogous to \edx, we first need to
recall some basic facts about the domain of integration.
The space of metrics on a compact topological surface $\Sigma$ modulo
diffeomorphisms and Weyl transformations is a finite dimensional compact space
$\CM_h$, known as moduli space. (It is 0-dimensional for genus $h=0$;
2-dimensional for $h=1$; and ($6h-6$)-dimensional for $h\ge2$). If for each
point $\tau\in \CM_h$, we choose a representative metric $\hat g\dup_{ij}$,
then the orbits generated by the diffeomorphism and Weyl groups acting on $\hat
g\dup_{ij}$
generate the full space of metrics on $\Sigma$. Thus given the slice
$\hat g(\tau)$, any metric can be represented in the form
$$f^*g=\ee{\ph}\, \hat g(\tau)\ ,$$
where $f^*$ represents the action of the diffeomorphism $f: \Sigma\to\Sigma$.

Since the integrand of \epf\ is diffeomorphism invariant, the functional
integral would be infinite unless we formally divide out by the volume of orbit
of the diffeomorphism group. This is accomplished by gauge fixing to the slice
$\hat g(\tau)$; the Jacobian that enters can be represented in terms of
Fadeev-Popov ghosts, as familiar from the analogous procedure in gauge theory.
We parametrize an infinitesimal change in the metric as
$$\delta g\dup_{zz}=\grad z \xi_z\ ,
\qquad\delta g\dup_{\zbar \zbar}=\grad {\zbar}\xi_{\zbar}$$
(where for convenience we employ complex coordinates, and recall that the
components $g\dup_{z\zbar}=g^{\zbar z}$ are parametrized by $\ee{\ph}$). The
measure $\CD g$ at $\hat g(\tau)$ splits into an integration $[\d\tau]$ over
moduli, an integration $\CD\ph$ over the conformal factor, and an integration
$\CD\xi\,\CD\bar\xi$ over diffeomorphisms. The change of integration variables
$\CD\delta g\dup_{zz}\,\CD\delta g\dup_{\zbar\zbar}=(\det\!\grad z
\det\!\grad{\zbar})\,\CD\xi\,\CD\bar\xi$ introduces the Jacobian $\det\!\grad z
\det\!\grad{\zbar}$ for the change from $\delta g$ to $\xi$. The determinants
in turn can be represented as
\eqn\eFP{\eqalign{\det\!\grad z\,\det\!\grad{\zbar}
 &= \int\CD b\,\CD c\,\CD \bar b\,\CD\bar c\
 \e{-\int\d^2\xi\,\sqrt g\, b_{zz} \grad {\zbar} c^z
    -\int\d^2\xi\,\sqrt g\, b_{\zbar \zbar}\grad z c^\zbar}\cr
&\qquad\equiv\int\CD({\rm gh})\ \e{-\Sgh(b,c,\bar b,\bar c)}\ ,}}
where $\CD({\rm gh})\equiv \CD b\,\CD c\,\CD \bar b\,\CD\bar c$ is an
abbreviation for the measures associated to the ghosts $b,c,\bar b,\bar c$;
$b_{zz}$ is a holomorphic quadratic differential, and $c^z$ ($c^\zbar$)
is a holomorphic (anti-holomorphic) vector.

Finally, the ghost measure $\CD({\rm gh})$ is
not invariant under the conformal transformation $g\to\ee\sigma g$,
instead we have \refs{\rpoly,\rFrlh,\rOA}\
\eqn\eagh{\CD_{e^\sigma\! g}({\rm gh})
=\e{-{26\over 48 \pi}\, S_L(\sigma,g)}\CD_{g}({\rm gh})\ ,}
where $S_L$ is again the Liouville action \eL.
(In units in which the contribution of a single scalar field to the
conformal anomaly is $c=1$, and hence $c=1/2$ for a single Majorana-Weyl
fermion, the conformal anomaly due to a spin $j$ reparametrization ghost
is given by $c=(-1)^F 2(1+6j(j-1))$. The contribution from a spin
$j=2$ reparametrization ghost is thus $c=-26$.)

We have thus far succeeded to reexpress the partition function \epf\ as
$$Z=\int [\d \tau]\ \CD_{g}\ph\ \CD_g({\rm gh})\ \CD_g X\
\e{-S_M-\Sgh-{\mu_0\over2\pi}\int \d^2\xi\,\sqrt g}\ .$$
Choosing a metric slice $g=\ee{\ph}\hat g$ gives
$$\CD_{e^{\ph}\hat g}\ph\,\CD_{e^{\ph}\hat g}({\rm gh})\,\CD_{e^{\ph} \hat g}X
= J(\ph,\hat g)\ \CD_{\hat g}\ph\,\CD_{\hat g}({\rm gh})\,\CD_{\hat g}X\ ,$$
where the Jacobian $J(\ph,\hat g)$ is easily calculated for the matter and
ghost sectors $\bigl($\edx\ and \eagh$\bigr)$ but not for the Liouville mode
$\ph$.
The functional integral over $\ph$ is complicated by the implicit metric
dependence in the norm
$$\|\delta \ph\|^2_g=\int \d^2\xi\,\sqrt g\,(\delta \ph)^2
=\int \d^2\xi\,\sqrt {\hat g}\, \ee{\ph}\, (\delta \ph)^2\ ,$$
since only if the $\ee{\ph}$ factor were absent above would the
$\CD_{\hat g}\ph$ measure reduce to that of a free field.

In \rDDK, it is simply {\it assumed\/}\foot{Some attempts to
justify this assumption may be found in \rMMHK.} that the overall Jacobian
$J(\ph,\hat g)$ takes the form of an exponential of a local Liouville-like
action
$\int \d^2\xi\,\sqrt {\hat g}\,(\tilde a\,\hat g^{ab} \del_a \ph \del_b \ph
+\tilde b \hat R \ph + \mu \ee{\tilde c \ph})$, where $\tilde a$, $\tilde b$,
and $\tilde c$ are constants that will be determined by requiring overall
conformal invariance ($\tilde c$ is inserted in anticipation of rescaling of
$\ph$). With this assumption, the partition function \epf\ takes the form
\eqn\epddk{\eqalign{Z=\int [\d \tau]\,\CD_{\hat g}\ph\,
\CD_{\hat g}({\rm gh})\,\CD_{\hat g}X\ &\e{-S_M(X,\hat g)
-\Sgh(b,c,\bar b,\bar c;\,\hat g)}\cr
&\qquad\cdot
\e{-\int \d^2\xi\,\sqrt {\hat g}\,
(\tilde a\,\hat g^{ab} \del_a \ph \del_b \ph
+\tilde b \hat R \ph + \mu \ee{\tilde c \ph})}\cr}}
where the $\ph$ measure is now that of a free field.

The path integral \epddk\ was defined to be reparametrization invariant, and
should depend only on $\ee{\ph}\hat g=g$ (up to diffeomorphism), not on the
specific slice $\hat g$.  Due to diffeomorphism invariance, \epddk\
should thus be invariant under the infinitesimal transformation
\eqn\eir{\delta \hat g=\varepsilon(\xi)\hat g\ ,\quad
\delta\ph=-\varepsilon(\xi)\ ,}
and we can use the known conformal anomalies \edx\ and \eagh\ for
$\ph$, $X$, and the ghosts to determine the constants
$\tilde a,\tilde b,\tilde c$.
Substituting the variations \eir\ in \epddk, we find terms of the form
$$\Bigl({D-26+1\over 48 \pi}+\tilde b\Bigr)\int \d^2\xi\,\sqrt {\hat g}\
\hat R\,\varepsilon\qquad{\rm and}\qquad
(2\tilde a-\tilde b) \int \d^2\xi\,\sqrt {\hat g}\ \varepsilon \lapl\ph\ ,$$
where the $D-26$ on the left is the contribution from the matter and ghost
measures $\CD_{\hat g}X$ and $\CD_{\hat g}({\rm gh})$,
and the additional 1 comes from the $\CD_{\hat g}\ph$ measure.
Invariance under \eir\ thus determines
\eqn\eab{\tilde b={25-D\over48 \pi}\ ,\quad \tilde a=\half \tilde b\ .}

Substituting the values of $\tilde a,\tilde b$ into the Liouville action in
\epddk\ gives
\eqn\eabs{{1\over 8 \pi}\int \d^2\xi\,\sqrt {\hat g}\,
\Bigl({25-D\over12}\hat g^{ab}\,\del_a\ph\,\del_b\ph
+{25-D\over6}\hat R\,\ph\Bigr)\ .}
To obtain a conventionally normalized kinetic term
${1\over8 \pi}\int (\del\ph)^2$, we rescale
$\ph\to \sqrt{{12\over25-D}}\,\ph$. (This normalization leads to the leading
short distance expansion $\ph(z)\,\ph(w)\sim -\log(z-w)$.)
In terms of the rescaled $\ph$, we write the Liouville action as
\eqn\ersL{{1\over 8 \pi}\int \d^2\xi\,\sqrt {\hat g}\,
\Bigl(\hat g^{ab}\,\del_a\ph\,\del_b\ph+Q\hat R\,\ph\Bigr)\ ,}
where
\eqn\eQ{Q\equiv\sqrt{{25-D\over3}}\ .}
The energy-momentum tensor
$T=-\half\del\ph\del\ph+{Q\over2}\del^2\ph$ derived from
\ersL\ has leading short distance expansion
$T(z)T(w)\sim {\ha c_{\rm Liouville}/(z-w)^4}+\ldots$,
where $c_{\rm Liouville}=1+3Q^2$. Note that if we substitute \eQ\
into $c_{\rm Liouville}$ and add an additional $c=D-26$ from the matter and
ghost sectors, we find that the
total conformal anomaly vanishes (consistent with the required overall
conformal invariance).

It remains to determine the coefficient $\tilde c$ in \epddk. We have since
rescaled $\ph$, so we write instead $\ee{\alpha\ph}$ and determine
$\alpha$ by the requirement that the
physical metric be $g={\hat g}\,\ee{\alpha\ph}$. Geometrically, this means that
the area of the surface is represented by
$\int \d^2\xi\,\sqrt {\hat g}\,\ee{\alpha\ph}$.
$\alpha$ is thereby determined by the requirement that $\ee{\alpha\ph}$ behave
as a (1,1) conformal field (so
that the combination $\d^2\xi\,\ee{\alpha\ph}$ is conformally invariant). For
the energy-momentum tensor
mentioned after \eQ, the conformal weight\foot{Recall that $h$ is given by the
leading term in the operator product expansion $T(z)\,\ee{\alpha\ph(w)}\sim
{h\,\ee{\alpha\ph}/(z-w)^2}+\ldots\ $. Recall also that for a conventional
energy-momentum tensor $T=-\half\del\ph\del\ph$, the conformal weight of
$\ee{ip\ph}$ is $h=\overline h=p^2/2$.} of $\ee{\alpha\ph}$ is
\eqn\ecw{h(\ee{\alpha\ph})
=\overline h(\ee{\alpha\ph})=-\half\alpha(\alpha-Q)\ .}
Requiring that $h(\ee{\alpha\ph})=\overline h(\ee{\alpha\ph})=1$
determines that $Q={2/\alpha}+\alpha$. Using \eQ\ and solving for $\alpha$
then gives\foot{The choice of root for $\alpha$ is determined by making
contact with the classical limit of the Liouville action. Note that the
effective coupling in \eabs\ goes as $(25-D)\inv$ so the classical limit
is given by $D\to-\infty$. In this limit the above choice of root has the
classical $\alpha\to0$ behavior.}
\eqn\ealph{\alpha={1\over\sqrt{12}}\bigl(\sqrt{25-D}-\sqrt{1-D}\bigr)\ .}

For spacetime embedding dimension $d\le1$, we find from \eQ\ and \ealph\
that $Q$ and $\alpha$ are both real (with $\alpha\le Q/2$). The $D\le1$
domain is thus where the Liouville theory is well-defined and most easily
interpreted. For $D\ge25$, on the other hand, both $\alpha$ and $Q$ are
imaginary. To define a real physical metric
$g=\ee{\alpha\ph}{\hat g}$, we need to Wick rotate $\ph\to-i\ph$. (This changes
the sign of the kinetic term for $\ph$. Precisely at $D=25$ we can interpret
$X^0=-i\ph$ as a free time coordinate. In other words, for a string naively
embedded in 25 flat euclidean dimensions, the Liouville mode turns out to
provide automatically a single timelike dimension,
dynamically realizing a string embedded in 26 dimensional minkowski spacetime.
Finally, in the regime $1<D<25$, $\alpha$ is complex, and $Q$ is imaginary.
Sadly, it is not yet known how to make sense of the Liouville approach for the
regime of most physical interest. We mention as well that so-called
non-critical strings (i.e.\ whose conformal anomaly is compensated by a
Liouville mode) in $D$ dimensions can always be reinterpreted as critical
strings in $D+1$ dimensions, where the Liouville mode provides the additional
(interacting) dimension. (The converse, however, is not true since it is not
always possible to gauge-fix a critical string and artificially disentangle the
Liouville mode.)

It remains to extract the string susceptibility $\gamma$ of
\epbl\ in this formalism.
We write the partition function for fixed area $A$ as
\eqn\eZA{Z(A)=\int\CD\ph\,\CD X\ \e{-S}\
\delta\Bigl({\textstyle\int} \d^2\xi\,\sqrt {\hat g}
\,\ee{\alpha\ph} - A\Bigr)\ ,}
where for convenience we now group the ghost determinant and integration over
moduli into $\CD X$. We define a string susceptibility $\gamma$ as in \elolnhg\
by
$$Z(A)\sim A^{(\gamma-2){\chi/2}-1}\ ,\quad A\to\infty\ ,$$
and determine $\gamma$ by a simple scaling argument. (Note that for genus zero,
we have $Z(A)\sim A^{\gamma-3}$ as in \eloln.)
Under the shift $\ph\to\ph+\rho/\alpha$ for $\rho$ constant, the measure in
\eZA\ does not change. The change in the action \ersL\ comes from the term
$${Q\over 8 \pi}\int \d^2\xi\,\sqrt {\hat g}\, \hat R\,\ph\to
{Q\over 8 \pi}\int \d^2\xi\,\sqrt {\hat g}\, \hat R\,\ph +
{Q\over 8 \pi}{\rho\over\alpha}\int \d^2\xi\,\sqrt {\hat g} \hat R\ .$$
Substituting in \eZA\ and using the Gauss-Bonnet formula
${1\over4 \pi}\int \d^2\xi\,\sqrt {\hat g} \hat R=\chi$ together with the
identity $\delta(\lambda x)=\delta(x)/|\lambda|$ gives
$Z(A)=\ee{-{Q \rho \chi/ 2\alpha}-\rho}\,Z(\ee{-\rho}A)$.
We may now choose $\ee\rho=A$, which results in
$$Z(A)=A^{-Q\chi/2\alpha-1}\,Z(1)= 
A^{(\gamma-2){\chi/2}-1}\,Z(1)\ ,$$
and we confirm from \eQ\ and \ealph\ that
\eqn\egam{\gamma=2-{Q\over \alpha}
={1\over12}\bigl(D-1-\sqrt{(D-25)(D-1)}\,\bigr)\ .}
This result reproduces \epbl, which we used to compare with the result of the
matrix model calculation (recall that $\gamma=-1/m$ for $D=1-6/m(m+1)\,)$.

\subsec{Dressed operators / dimensions of fields}

Now we wish to determine the effective dimension of fields after coupling
to gravity. Suppose that $\Phi_0$ is some spinless primary field in a
conformal field theory with conformal weight $h_0=h(\Phi_0)=\overline
h(\Phi_0)$ before coupling to gravity. The gravitational ``dressing'' can
be viewed as a form of wave function renormalization that allows $\Phi_0$
to couple to gravity. The dressed operator $\Phi=\ee{\beta\ph}\Phi_0$ is
required to have dimension (1,1)
so that it can be integrated over the surface $\Sigma$ without breaking
conformal invariance. (This is the same argument used prior to \ealph\ to
determine $\alpha$). Recalling the formula \ecw\ for the conformal weight of
$\ee{\beta\ph}$, we find that $\beta$ is determined by the condition
\eqn\ehdr{h_0-\half \beta(\beta-Q)=1\ .}

We may now associate a critical exponent $h$ to the behavior of the one-point
function of $\Phi$ at fixed area $A$,
\eqn\eopt{F_\Phi(A)\equiv {1\over Z(A)}\int\CD\ph\,\CD X\ \e{-S}\
\delta\Bigl({\textstyle\int} \d^2\xi\,\sqrt {\hat g}\,\ee{\alpha\ph} - A\Bigr)
\ {\textstyle\int} \d^2\xi\,\sqrt{\hat g}\, \ee{\beta\ph}\,\Phi_0
\sim A^{1-h}\ .}
This definition conforms to the standard convention that
$h<1$ corresponds to a relevant operator, $h=1$ to a marginal operator, and
$h>1$ to an irrelevant operator (and in particular that relevant operators tend
to dominate in the infrared, i.e.\ large area, limit).

To determine $h$, we employ the same scaling argument that led to \egam. We
shift $\ph\to\ph+\rho/\alpha$ with $\ee\rho=A$ on the right hand side of \eopt,
to find
$$F_\Phi(A)
={A^{-Q\chi/2\alpha-1+\beta/\alpha}\over A^{-Q\chi/2\alpha-1}}\,F_\Phi(1)
=A^{\beta/\alpha}\,F_\Phi(1)\ ,$$
where the additional factor of $\ee{\rho\beta/\alpha}=A^{\beta/\alpha}$ comes
from the $\ee{\beta \phi}$ gravitational dressing of $\Phi_0$.
The gravitational scaling dimension $h$ defined in \eopt\ thus satisfies
\eqn\ehba{h=1-\beta/\alpha\ .}
Appealing to the semiclassical argument employed before \ealph,
we solve \ehdr\ for $\beta$ with the branch
$$\beta=\half Q-\sqrt{{\textstyle {1\over4}}Q^2-2+2h_0}
={1\over\sqrt{12}}\bigl(\sqrt{25-D}-\sqrt{1-D+24 h_0}\bigr)$$
(for which $\beta\le Q/2$, and $\beta\to0$ as $D\to-\infty$).
Finally we substitute the above result for $\beta$ and the value \ealph\ for
$\alpha$ into \ehba, and find\foot{We can also substitute $\beta=\alpha(1-h)$
from \ehba\ into \ehdr\ and use $-\half\alpha(\alpha-Q)=1$ (from before \ealph)
to rederive the result $h-h_0=h(1-h){\alpha^2/2}$ for
the difference between the ``dressed weight'' $h$ and the bare weight $h_0$
\rKPZ.}
\eqn\ehf{h={\sqrt{1-D+24h_0}-\sqrt{1-D}\over\sqrt{25-D}-\sqrt{1-D}}\ .}

As an example, we apply these results to the minimal models \rBPZ\ mentioned at
the end of section 2. These have a set of operators labelled by two integers
$p,q$ (satisfying $1\le r\le q-1,\ 1\le s\le p-1$). Coupled to gravity, these
operators turn out to have dressed conformal weights
\eqn\edrs{h_{r,s}={p+q-|pr-qs|\over p+q-1}
\qquad 1\le r\le q-1,\ 1\le s\le p-1\ ,}
in agreement with the weights determined from the $(p,q)$ formalism discussed
in section 3 for the generalized KdV hierarchy (see e.g.\ \refs{\rD,\rGGPZ}).

More explicitly, we consider
the first member of the unitary discrete series, i.e.\ the $D=1/2$ Ising model,
which has $(p,q)=(4,3)$. Before coupling to gravity, critical exponents
$\nu,\alpha,\beta$ can be defined in terms of the divergences of correlation
length $\xi\sim t^{-\nu}$, specific heat $C\sim t^{-\alpha}$, and magnetization
$m\sim t^\beta$ with respect to the deviation $t=(T-T_c)/T_c$ from the critical
temperature $T_c$. In terms of the conformal weights of the energy and spin
operators $h_\varepsilon$ and $h_s$, these exponents satisfy
$\nu={1\over 2(1-h_\varepsilon)}$, $\alpha=2(1-\nu)$,
$\beta=(2-\alpha)h_s$. According to \edrs, the coupling to gravity
induces the shifts $h_\varepsilon=\half\to{2\over3}$,
$h_s={1\over16}\to{1\over6}$, which implies corresponding shifts in
$\nu$, $\alpha$, and $\beta$.

\listrefs\bye